\documentclass[fleqn,usenatbib]{mnras}

\usepackage{amssymb}
\usepackage{xcolor,subcaption}
\usepackage[export]{adjustbox}
\usepackage{flushend}
\usepackage[utf8]{inputenc}

\usepackage[compact]{titlesec}  
\titlespacing{\section}{0pt}{12pt}{7pt}
\titlespacing{\subsection}{0pt}{9pt}{4pt}

\def\comment#1{#1}

\newcommand{\frb}{FRB~20180916B}
\newcommand{\psv}{\texttt{PI}}
\newcommand{\ptt}{\texttt{PII}}

\newcommand{\ru}{\,hr$^{-1}$}

\newcommand\eff{Effelsberg}
\newcommand\rss{FRB~20201124A}

\newcommand{\rAO}{FRB~20121102A}

\usepackage[T1]{fontenc}

\DeclareRobustCommand{\VAN}[3]{#2}
\let\VANthebibliography\thebibliography
\def\thebibliography{\DeclareRobustCommand{\VAN}[3]{##3}\VANthebibliography}

\usepackage{graphicx}	
\usepackage{subcaption} 
\usepackage{booktabs}   
\usepackage{amsmath}	
\usepackage{amssymb}	
\usepackage{newtxtext,newtxmath}
\usepackage[switch,columnwise]{lineno}

\defcitealias{21PastorR3}{PM21}
\defcitealias{21ZiggyR3}{ZP21}
\defcitealias{21DongziPA}{DL21}
\graphicspath{{figures/}}

\title[\frb\ at 5GHz]{High frequency study of \frb~using the 100-m \eff~radio telescope}

\author[Bethapudi et al.]{S. Bethapudi$^1\thanks{Email: \href{mailto:sbethapudi@mpifr-bonn.mpg.de}{sbethapudi@mpifr-bonn.mpg.de}}$,
L.~G.~Spitler$^{1}$,
R.~A.~Main$^{1}$,
D.~Z.~Li$^{2}$,
R.~S.~Wharton$^{3}$. \\ 
$^{1}$Max-Planck-Institut f{\"u}r Radioastronomie, Auf dem H{\"u}gel 69, D-53121 Bonn, Germany \\
$^{2}$Cahill Center for Astronomy and Astrophysics, MC 249-17 California Institute of Technology, Pasadena CA 91125, USA \\
$^{3}$NASA Postdoctoral Program Fellow, Jet Propulsion Laboratory, California Institute of Technology, Pasadena, CA 91109, USA 
}

\date{Accepted XXX. Received YYY; in original form ZZZ}

\pubyear{2022}

\begin{document}
\label{firstpage}
\pagerange{\pageref{firstpage}--\pageref{lastpage}}
\maketitle

\begin{abstract}

\par \frb~is a repeating fast radio burst (FRB) with an activity period of $16.33$ days. 
In previous observations ranging from $\sim 150-1400$\,MHz, the activity window was found to be frequency dependent, with lower frequency bursts occurring later. In this work, we present the highest-frequency detections of bursts from this FRB, using the 100-m Effelsberg Radio Telescope at 4$-$8 GHz.
We present the results from two observing campaigns. We performed the first campaign over an entire activity period which resulted in no detections. The second campaign was in an active window at 4$-$8\,GHz which we predicted from our modelling of chromaticity, resulting in eight burst detections.
The bursts were detected in a window of 1.35 days, 3.6 days preceding the activity peak seen by CHIME, suggesting the chromaticity extends to higher frequency.
The detected bursts have narrower temporal widths and larger spectral widths compared to lower frequencies. All of them have flat polarization position angle sweeps and high polarization fractions. The bursts also exhibit diffractive scintillation due to the Milky Way, following a $f^{3.90\pm0.05}$ scaling, and vary significantly over time.
We find that burst rate across frequency scales as $f^{-2.6\pm0.2}$.
Lastly, we examine implications of the frequency dependency on the source models.

\end{abstract}

\begin{keywords}
methods: observational -- techniques: miscellaneous -- transients: fast radio bursts -- scattering
\end{keywords}



\section{Introduction}

\par Fast Radio Bursts (FRBs) are bright, microseconds- to milliseconds-duration transients observed from hundreds of MHz to several GHz and which originate from extragalactic distances. 
The first FRB was discovered in $2007$ \citep{lorimer2007}, and the lack of repeatability motivated theoretical models of cataclysmic origins \citep{Thornton+13}. 
Discovery of repeat bursts from \rAO~\citep{Spitler2016} brought into light a new class of repeating FRBs. 
The focus of the paper is one such repeating FRB, \frb, which was discovered by Canadian Hydrogen Intensity Mapping Experiment \citep[CHIME]{chime}. 

\par \frb~is a repeating FRB with an activity period of $\sim 16$ days and active window of $\sim 5$ days at $600$ MHz~\citep{20R3Period}. 
A well determined periodicity meant that observatories know when to observe to increase the likelihood of detecting bursts. \citet{20Marcote}~provided milli-arcsecond localization of the source by means of Very Long Baseline Interferometry (VLBI) with the European VLBI Network (EVN) associating the FRB to a star-forming region of a massive spiral galaxy at redshift of 0.0337. 
\citet{tendulkar2021}~and \citet{mannings+21}~studied the local region with the Hubble Space Telescope (HST) and Gran Telescopio Canarias (GTC), from which \citet{tendulkar2021}~postulated that \frb~is likely to be either an old neutron star/magnetar High Mass X-ray Binary (HMXBs) or $\gamma$-ray binary with late OB-type star. 
Multiple X-ray campaigns 
have already been performed, however no contemporaneous signal has been detected yet \citep{20ScholzR3Xray,20TavaniGX, 20PiliaR3}. 
Detection of prompt/contemporaneous emission at different wavelengths when an FRB is detected helps place direct limits on the type of emission mechanisms and nature of the source. Moreover, presence of any persistent emission in different wavelengths helps in studying the local environment in much greater detail. 


\par \frb~has been detected by multiple instruments at different radio frequencies across different cycles. 
LOFAR has twenty seven at $150$ MHz \citep{21PastorR3,21ZiggyR3}, GMRT five at $300$ MHz \citep{21ZiggyR3}, GBT eight at $300$ MHz \citep{20Chawla}, SRT three at $328$ MHz \citep{20PiliaR3}, GMRT four at $400$ MHz \citep{21SandR3}, CHIME/FRB fifty five at $600$ MHz \citep{20R3Period,21ZiggyR3}, GMRT fifteen at $650$ MHz \citep{marthi+20}, GBT seven at $800$ MHz \citep{21SandR3}, VLA/realfast one at $1350$ MHz \citep{AtelRF}, APERTIF fifty four at $1370$ MHz \citep{21PastorR3}, and lastly \eff~(as part of European VLBI Network) four at $1700$ MHz \citep{20Marcote}. Note that all of these detections are at comparatively low-frequency ($\lesssim 2$ GHz). 

\par The idea of observing the source at high frequencies is not novel.
\citet{20PearlmanDSN} performed a hundred hour observing campaign on \frb~using the Deep Space Network (DSN) in $S-$ ($2.3$ GHz) and $X-$ ($8.4$ GHz) bands simultaneously but did not detect any bursts. 
\rAO~has been observed up to 8 GHz \citep{18GajjarR1} using the GBT. FRB~20190520 \citep{Niu+21} has also been studied with the GBT at $6$ GHz \citep{AnnaThomas+22}. 
FRB~20200120E \citep{Bhardwaj2021} was studied with the DSN in $\sim 2$ GHz \citep{21MajidM81}. 
High frequency detections are particularly useful in measuring RMs. For example, in case of \rAO{} and FRB~20190520, due to the extremely high RM, measuring it at higher frequencies is much easier \citep{Hilmarsson2021, AnnaThomas+22}.

\par \frb{} burst detections by LOFAR and APERTIF \citep[hereafter ZP21, PM21]{21ZiggyR3,21PastorR3} revealed the frequency dependency in the occurrence of bursts. Specifically, it was noted that bursts at lower frequencies occur later and bursts at higher frequencies occur earlier. 
In other words, the \frb{} source not only exhibited periodicity, but also ``chromaticity'' since bursts showed selectivity in phase. In this work, the proposition that the chromaticity continues to higher frequencies is tested.  

\par The structure of the paper is as follows: Sec.~\ref{sec:obs}~explains the strategy involved in scheduling observations and describes the search procedures.
Sec.~\ref{sec:det}~presents all the detected bursts, their polarization, scintillation and flux-fluence properties. A rate estimate is also reported using literature rates at various frequencies, and the rate-frequency relation is modeled.  Sec.~\ref{sec:dis}~presents various discussions of the results presented here. 
Lastly, conclusions are presented in Sec.~\ref{sec:con}.

\section{Observations and searches} \label{sec:obs}

\par This section is organized as follows: we describe the observatory and the backend used in Sec.~\ref{ssec:eff}, construct a chromaticity model in Sec.~\ref{ssec:chrom}, report the observations performed in Sec.~\ref{ssec:obs}, and finally describe the search strategy employed in Sec.~\ref{ssec:search}. 

\subsection{Effelsberg} \label{ssec:eff}

\par Observations were conducted using the 100-m \eff~Radio Telescope using the S45mm receiver which has linearly polarized feeds and a System Equivalent Flux Density (SEFD) of 25 Jy. The receiver is connected to the \texttt{ROACH2} backend, which provides \texttt{8bit} full Stokes filterbank data at $131.072~\mu s$ time resolution in the $4-8$ GHz band.  
It records the entire band as two separate sub-bands, which are $4-6$ GHz and $6-8$ GHz, each channelized into $2048$ channels.
\eff~observes strong system artefacts in the $4-8$ GHz band which reduces sensitivity to bursts, in particular below $4.5$ GHz. 

\subsection{Chromaticity model} \label{ssec:chrom}

\begin{table}
\centering
\caption{Parameters for the power laws presented in this work. The basic form of the power law over frequency $f$ considered here is $B \big( f / 600 {\rm\ MHz} \big)^A$. Errors in Burst Rate are 95\%. All other errors are $1\sigma$.  Units are provided wherever relevant. }
\label{tab:pl}
\begin{tabular}{@{}cccc@{}}
\toprule
Equation & B  & A & Refer to  \\ \midrule
\begin{tabular}[c]{@{}c@{}}Peak activity\\ phase ($\phi_P$)\end{tabular} & 0.47 $\pm$ 0.02 & -0.23 $\pm$ 0.05 & Sec.~\ref{ssec:chrom} \\
\begin{tabular}[c]{@{}c@{}}FWHM phase\\ ($\Delta_P$)\end{tabular} & 53.4 hr $\pm$ 15.6 hr & -0.35 $\pm$ 0.32 & Sec.~\ref{ssec:chrom} \\
Burst Rate & 22.8 $\pm$ 4.7 \ru & -2.6 $\pm$ 0.2 & Sec.~\ref{ssec:rate} \\
\begin{tabular}[c]{@{}c@{}}Scintillation\\ bandwidth\end{tabular} & 0.82 $\pm$ 0.06 kHz & 3.90 $\pm$ 0.05 & Sec.~\ref{ssec:scint} \\ \bottomrule
\end{tabular}
\end{table}

\par
\begin{figure}
\includegraphics[width=0.5\textwidth,keepaspectratio]{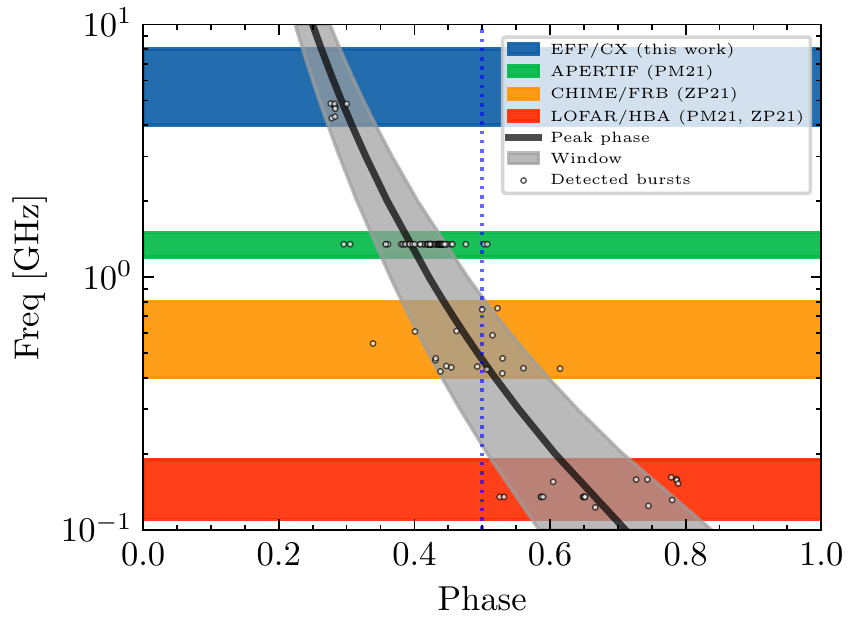}
\caption{Frequency dependency on the activity phase. The black bold line shows peak-phase frequency dependency ($\phi_P$), the shaded grey region is the FWHM ($\Delta_P$). Colored bands are observing frequency ranges of various observatories. White dots in each band are the detections by that particular observatory. The vertical dotted blue line denotes $\phi=0.5$. Higher frequency bursts arrive at earlier phase and active window shortens at higher frequencies. Sec.~\ref{ssec:chrom} describes the steps taken to produce this fit.
}
\label{fig:chrom}
\end{figure}

 \citetalias{21ZiggyR3}~provides the most up-to-date periodicity model with the reference ${\rm MJD} = 58369.40$ and period of $16.33$ days at $600$ MHz. The reference MJD corresponds to the start of the first activity cycle in which the source was discovered. 
This periodicity model is designed to have bursts arriving at $600$ MHz (CHIME band) at $\phi=0.5$ (see Fig.~\ref{fig:chrom}, c.f. \citetalias[Fig. 4]{21PastorR3}, \citetalias[Fig. 9]{21ZiggyR3}).  

\par We construct a chromatic model as a power-law relation between observing frequency and detected burst phase. Using this model, we predict active windows at $4-8$ GHz. The burst detections used to construct our model are chosen carefully: since bursts detected in follow-up campaigns might introduce biases into the model, only a subset of all the detections is used, which are APERTIF L-band detections reported in \citetalias{21PastorR3}, CHIME/FRB detections reported in \citetalias{21ZiggyR3}, and LOFAR detections reported in \citetalias{21PastorR3,21ZiggyR3}. 
We chose only observations that have performed a blind search over the entire window, and therefore is not biased by any observing choice.
Given that the frequency extent of the bursts is varying within an observing bandwidth, the frequency of the bursts is computed as the mean of the start/stop frequencies of the burst. Bursts where the start/stop frequencies are not reported are simply excluded. 

\par Then, bursts are binned into frequency bands of $100$ MHz each starting from $100$ MHz to $2$ GHz. For each band, the mean and standard deviation of the burst phases (which are computed by folding at the activity period) in the frequency band are computed. The mean of the phase is treated as peak activity phase $(\phi_P)$ and the standard deviation is used to compute the full-width-at-half-maximum (FWHM, $\Delta_P$). 
The mean of the burst frequencies in a band is treated as the frequency of the band instead of using the center frequency of the band.

\par Now, power laws are fitted for $\phi_P$ and $\Delta_P$ against frequency $f$, of the form 
\begin{equation} \label{eq:plform}
    B~\bigg( \frac{f}{600 {\rm~MHz}} \bigg)^A,
\end{equation}
where $A$ and $B$ are fit parameters. The fitting is done using \texttt{scipy.optimize.curve\_fit}. The fitting yields peak phase and FWHM parameters shown in Tab.~\ref{tab:pl}. The errors are $1\sigma$.

\par The peak activity phase and FWHM at any frequency $f$ is directly given by $\phi_P$, $\Delta_P$, where the start and stop phases of the active window are computed as $\phi_P - \frac{1}{2}\Delta_P$ and $\phi_P + \frac{1}{2}\Delta_P$. 
The power-law is graphically presented in Fig.~\ref{fig:chrom}. 
The fitted power laws are extended to the bandwidth of interest ($4-8$ GHz) and specific phase regions are used to predict time windows to schedule observations in. 

\par 
The chromatic model fitted (Tab.~\ref{tab:pl}) is not a periodic model and therefore has limitations. Since it is a powerlaw, it asymptotes to zero at sufficiently high frequencies and cannot wrap around,
making inferences at high frequencies difficult.
Nevertheless, we begin with the assumption that it is valid over the frequency range of interest ($4-8$ GHz).

\subsection{Observations}
\label{ssec:obs}

\par Each observation started with a test pulsar scan (PSR~B0355+54) and a noise-diode scan to check data quality and to provide polarization calibration solution, respectively.
In the Cycle 65 observation the $6-8$ GHz subband recording failed, hence only the lower subband ($4-6$ GHz) was recorded.

\begin{table}
\centering
\caption{4-8 GHz observations with Effelsberg. Cycle refers to the activity cycle since the reference MJD. Exposure is the total time spent on \frb~in hours. The start time is provided in UTC. Phase range corresponds to the exposure in phase units. The number of bursts detected are written in the Bursts column.}
\begin{tabular}{@{}ccccc@{}}
\toprule
Cycle & \begin{tabular}[c]{@{}l@{}}Exposure\\ {[}hr{]}\end{tabular} & Start time & Phase range & Bursts \\ \midrule
42 & 2.22 & 2020-08-09T16:18:15 & 0.944-0.950 & 0 \\
43 & 0.26 & 2020-08-10T19:17:35 & 0.013-0.014 & 0 \\
43 & 1.17 & 2020-08-11T04:25:09 & 0.036-0.039 & 0 \\
43 & 0.96 & 2020-08-12T05:05:56 & 0.099-0.102 & 0 \\
43 & 2.00 & 2020-08-13T04:07:52 & 0.158-0.163 & 0 \\
43 & 1.00 & 2020-08-14T04:25:13 & 0.220-0.223 & 0 \\
43 & 1.00 & 2020-08-15T20:32:28 & 0.322-0.325 & 0 \\
43 & 0.94 & 2020-08-17T04:34:38 & 0.404-0.407 & 0 \\
43 & 2.26 & 2020-08-18T03:40:48 & 0.463-0.469 & 0 \\
43 & 2.55 & 2020-08-20T00:06:37 & 0.577-0.583 & 0 \\
43 & 2.19 & 2020-08-21T03:59:26 & 0.648-0.653 & 0 \\
43 & 1.82 & 2020-08-23T06:11:59 & 0.776-0.780 & 0 \\
43 & 1.78 & 2020-08-24T03:52:48 & 0.831-0.836 & 0 \\
43 & 1.81 & 2020-08-25T04:04:49 & 0.893-0.897 & 0 \\
44 & 2.36 & 2020-08-27T03:36:43 & 0.014-0.020 & 0 \\
44 & 1.85 & 2020-08-28T03:42:22 & 0.076-0.080 & 0 \\
44 & 1.52 & 2020-08-29T03:30:17 & 0.136-0.140 & 0 \\
44 & 4.20 & 2020-08-30T10:14:19 & 0.215-0.225 & 0 \\
44 & 2.02 & 2020-09-01T03:29:26 & 0.320-0.325 & 0 \\
44 & 1.77 & 2020-09-03T03:54:42 & 0.444-0.448 & 0 \\ \midrule
60 & 5.23 & 2021-05-19T15:05:52 & 0.271-0.284 & 2 \\
61 & 5.13 & 2021-06-05T06:43:21 & 0.29-0.303  & 0 \\
62 & 4.50 & 2021-06-21T15:27:00 & 0.292-0.304 & 0 \\
65 & 3.32 & 2021-08-09T17:32:27 & 0.298-0.307 & 1 \\
67 & 3.35 & 2021-09-11T07:38:22 & 0.294-0.303 & 0 \\
68 & 3.99 & 2021-09-27T08:26:18 & 0.276-0.286 & 5 \\
69 & 3.99 & 2021-10-14T01:37:38 & 0.299-0.31  & 0 \\
70 & 2.99 & 2021-10-29T23:10:25 & 0.273-0.281 & 0 \\
71 & 5.99 & 2021-11-15T17:23:14 & 0.299-0.315 & 0 \\
72 & 5.49 & 2021-12-01T18:05:17 & 0.281-0.295 & 0 \\ \bottomrule
\end{tabular}%
\label{tab:obs}
\end{table}

\begin{figure}
\includegraphics[width=0.5\textwidth,keepaspectratio]{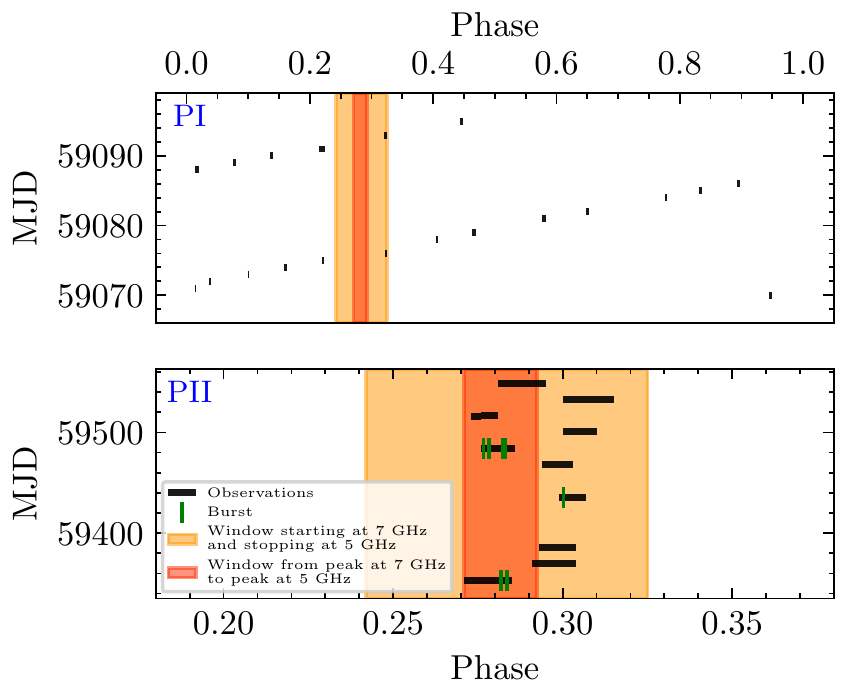}
\caption{Exposure plotted over MJD and phase for the two observing campaigns. \emph{Top}: \psv~campaign. \emph{Bottom}: \ptt~campaign. See Sec.~\ref{ssec:chrom} and Sec.~\ref{ssec:obs}. Solid black lines correspond to the exposure. Vertical green dashes denote detected bursts. Light orange region corresponds to a phase window starting at 7 GHz and stopping at 5 GHz, dark orange corresponds to phase window from peak at 7 GHz to peak at 5 GHz. }
\label{fig:mjdp}
\end{figure}

\par The initial observing strategy was to cover an entire period. These observations will be denoted by \psv. A total exposure of $35.73$ hours was distributed over $20$ days (covering an entire activity cycle of $16.33$ days) resulting an average of $1.78$ hours on source per day. This campaign did not lead to any detections.  

\par When the evidence for a chromatic window became stronger with \citetalias{21ZiggyR3} and \citetalias{21PastorR3}, the strategy was adapted by developing a chromatic model (Sec.~\ref{ssec:chrom}). 
To this end, $40$ hours was used to schedule observations in the predicted active windows. These will be denoted by \ptt.

Given the inherent uncertainties in the model, the observations were scheduled to be in a phase window starting at $7$ GHz and ending at $5$ GHz, which corresponds to a phase region $[0.238, 0.321]$ (c.f. Fig.~\ref{fig:chrom}, Peak phase, FWHM equations from Tab.~\ref{tab:pl}). 
\eff~records persistent, periodic radio frequency interference (RFI) below $4.5$ GHz,
hence we chose to focus on a window above 5 GHz.
To make the window symmetric, the start was chosen to be $7$ GHz, and the resulting length of this window is $32.5$ hours. 
Scheduling in the window was done according to the telescope availability.

\par All the observations performed, resulting exposures and number of bursts are tabulated in Tab.~\ref{tab:obs}. The same information is also presented graphically in Fig.~\ref{fig:mjdp}, where the black lines denote exposure against MJD and phase and the vertical green dashes are detected bursts.

\subsection{Search for bursts} \label{ssec:search}

\par All searches for bursts were performed using the \texttt{PRESTO} software package \citep{presto}.
The $4-6$ GHz and $6-8$ GHz sub-bands were searched separately (see Sec~\ref{ssec:eff}). Since the dispersion measure (DM) of the source is well known, single pulses are searched at only one DM of value of $348.820$  pc cm$^{-3}$ \citep{20R3Period}.
Any error in DM would not result in significant changes to detected bursts, as the observing frequency is quite high. 
\comment{The in band DM delay in $4-6$ GHz is about $50$ ms (corresponding to $\sim 400$ time samples). For the $6-8$ GHz, it is about $17$ ms ($\sim 130$ time samples). This delay is sufficient to accurately distinguish between RFI and potential candidates.}

\par Each filterbank file was re-processed, which involved whitening  every channel (subtracting mean, dividing by standard deviation and appropriately scaling into \texttt{uint8} bytes).
This step removed any bandshape (scales/offsets over channels) and narrow-band persistent RFI and effectively re-digitized the data. 
After re-digitizing the filterbank data, \texttt{PRESTO}'s \texttt{rfifind} tool was used to generate a mask to mitigate RFI. 
The mask was used to de-disperse and create a time series of the data.
\texttt{single\_pulse\_search.py} was run on the time series to extract all the single pulses candidates. The minimum S/N threshold used was $6$ and the \texttt{-b} flag was passed so not to exclude bad blocks. 
The maximum boxcar width was set to $50$ samples, which corresponds to $\sim 6.5$ ms. 
Candidates were plotted using custom \texttt{python} code and were manually vetted.

\section{Detections} \label{sec:det}

\begin{table*}
\centering
\caption{Table showing the measured properties of the eight bursts detected. MJD refers to the UTC Topocentric timestamp when de-dispersed to $5999.51171875$ MHz.
Phases of the bursts are reported in phase column. 
Peak flux and fluence are calculated from the calibrated time profile of the bursts. 
Widths are provided as upper limits wherever the bursts are unresolved. 
L-fraction is the total unbiased linear polarization fraction of the bursts.
Note that baselines from each of the Stokes time series profile has been removed which causes the unbiased linear polarization to exceed $100\%$.
V-fraction is the total circular polarization fraction of the bursts.
The frequency extent of the bursts is reported in Freq low and Freq high columns. Mean PPA is the Polarization Position Angle (PPA) averaged over the burst width.}
\label{tab:bursts}
\resizebox{\textwidth}{!}{%
\begin{tabular}{@{}ccccccccccc@{}}
\toprule
Burst & MJD & Phase & \begin{tabular}[c]{@{}l@{}}Peak flux\\ {[}Jy{]}\end{tabular} & \begin{tabular}[c]{@{}l@{}}Fluence\\ {[}Jy ms{]}\end{tabular} & \begin{tabular}[c]{@{}l@{}}Width\\ {[}ms{]}\end{tabular} & \begin{tabular}[c]{@{}l@{}}L-fraction\\ {[}\%{]}\end{tabular} & \begin{tabular}[c]{@{}l@{}}V-fraction\\ {[}\%{]}\end{tabular} & \begin{tabular}[c]{@{}l@{}}Freq low\\ {[}MHz{]}\end{tabular} & \begin{tabular}[c]{@{}l@{}}Freq high\\ {[}MHz{]}\end{tabular} & \begin{tabular}[c]{@{}l@{}}Mean PPA\\ {[}deg{]}\end{tabular} \\ \midrule
A & 59353.802118639 & 0.28 &0.44 & 0.08 & $\leq$ 0.26 & 72$\pm$14 & 33$\pm$16 & 4490.0 & 5075.2 & -56.2 \\
B & 59353.830062663 & 0.28 &0.75 & 1.49 & 3.42 & 105$\pm$3 & 13$\pm$3 & 4369.8 & 5041.0 & -62.3 \\
C & 59435.751975537 & 0.3 &0.54 & 0.09 & $\leq$0.26 & 80$\pm$12 & 15$\pm$13 & 4519.3 & 5182.7 & 58.3 \\
D & 59484.357407934 & 0.27 &1.82 & 0.42 & $\leq$0.26 & 142$\pm$5 & 1$\pm$3 & 4386.4 & 5348.8 & -61.4 \\
E & 59484.383221707 & 0.25 &0.94 & 0.14 & $\leq$0.26 & 105$\pm$7 & 2$\pm$5 & 4128.5 & 4391.3 & -48.9 \\
F & 59484.449512949 & 0.28 &0.36 & 0.13 & $\leq$0.53 & 101$\pm$12 & 22$\pm$10 & 4685.4 & 5018.6 & -61.8 \\
G & 59484.458202382 & 0.28 &1.84 & 0.46 & $\leq$0.26 & 111$\pm$3 & 7$\pm$2 & 4048.4 & 4586.7 & -61.0 \\
H & 59484.464427655 & 0.28 &0.18 & 0.12 & 0.79 & 101$\pm$15 & 20$\pm$12 & 4365.9 & 4914.0 & -72.6 \\ \bottomrule
\end{tabular}%
}
\end{table*}

\begin{figure*}\centering%
    \includegraphics[width=\textwidth, keepaspectratio]{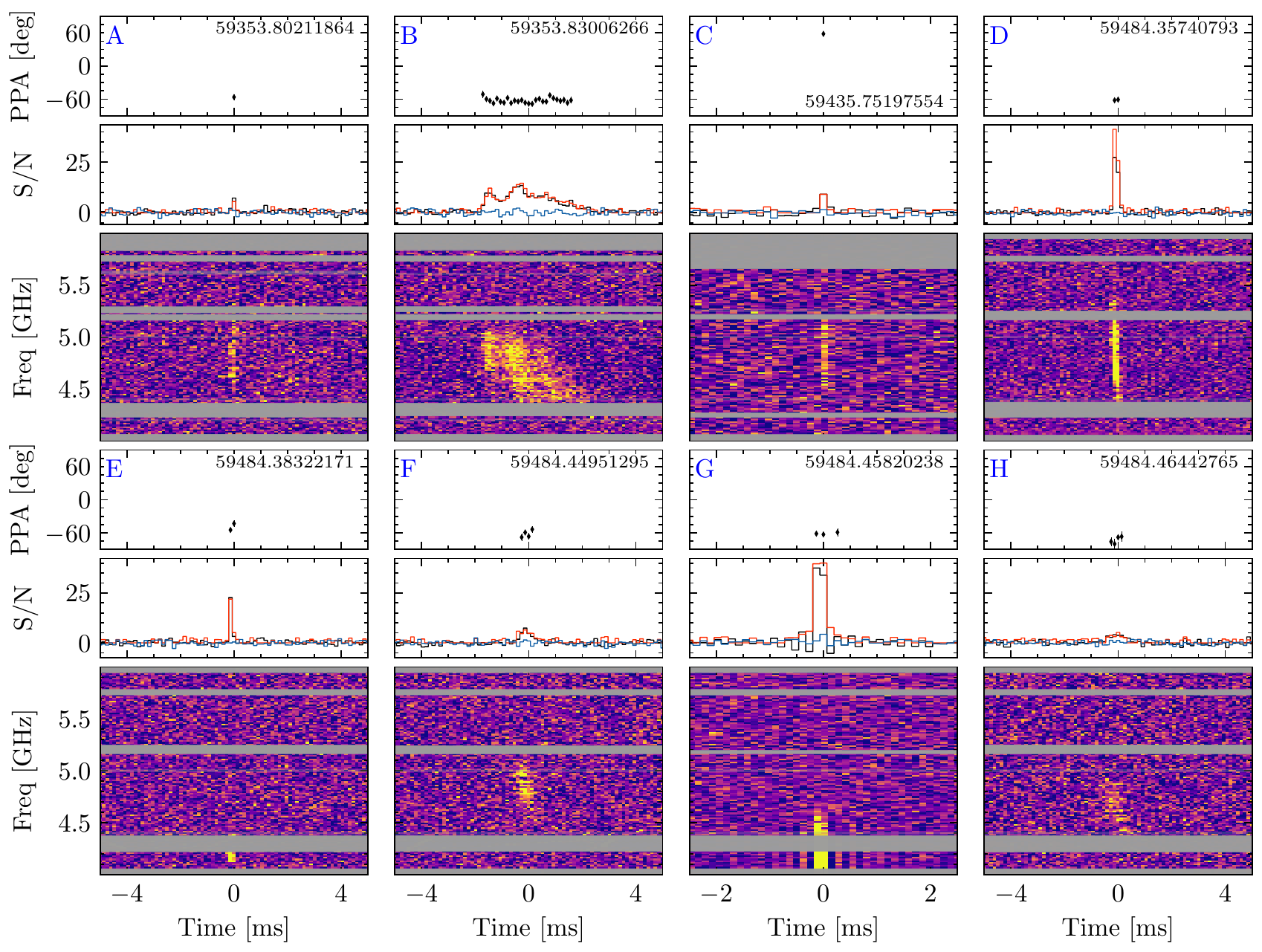}
    \caption{Portrait of the detected bursts. The \emph{bottom} panel shows de-dispersed filterbank with time/frequency resolution of $131.072~\mu$s and $15.625$ MHz. The \emph{middle} panel shows unbiased total intensity (black), linear (red), circular (blue) polarization time series in S/N units. The \emph{top} panel shows the polarization position angle sweep in degrees. The horizontal grey bands are masked RFI.}
    \label{fig:bursts}
\end{figure*}

\par We report detection of eight bursts:
two bursts detected in the first observation of the campaign (Cycle 60), one in Cycle 65, and five in a single observation (Cycle 68). All the detections reported here are only in the lower sub-band of $4-6$ GHz. 
The bursts are extracted from the recorded full-stokes data into fold-mode PSRFITS archives, and all subsequent analysis is performed on these archives. Care has been taken that the burst archives show the correct timestamp.
\par No attempt is made in estimating dispersion measure (DM) since the bursts are at high frequencies, possess narrow bandwidths, and the time resolution is not sufficient to resolve the burst structure in all of our bursts. Instead, the most accurate DM estimate, .viz $348.772$ pc cm$^{-3}$ \citep{21MicroR3} is used throughout the analysis as the DM. 

\par Time-of-Arrivals (TOAs) and phases are tabulated in Tab.~\ref{tab:bursts},  UTC topocentric and referenced to the center of the topmost channel ($5999.51171875$ MHz). Since the majority of the bursts are not time resolved, the \texttt{PRESTO} timestamp is used to calculate the TOA.
The spectral widths are visually measured by plotting, and reported in the table. 
All the bursts have a bandwidth of around $200$ MHz, $\sim 5\%$ of the observing bandwidth.
The time resolution of the data proved to be insufficient to resolve the burst structures of majority of the bursts. Hence, the widths reported are to be treated as upper limits for all but bursts-\texttt{B,H}. Moreover, for the same reason, the relationship between widths and observing frequency is not probed in detail.

\subsection{Polarization} \label{ssec:pol}

\par The S45mm receiver is equipped with a noise diode, which produces a square wave pure $U$, $100\%$ polarized calibration signal with a period of $1$ second and $50\%$ duty cycle 
. Two minutes of the noise diode scan was recorded at the start of every observation. 
The scans were folded at the diode period, and any RFI present was exhaustively flagged using \texttt{pazi} \citep{psrchive}.  These scans were then used to compute a calibration solution using \texttt{pac} from \texttt{PSRCHIVE} \citep{10wvs_pac,psrchive}, which was then applied to the burst archives. This step also performs the parallactic angle correction. 
The polarization calibration solution was validated by applying it to the test pulsar (B0355+54), which was observed at the start of every session. The calibrated polarization profile and the polarization position angle (PPA) sweep were consistent in all the sessions where bursts were detected and also was tested against reference pulse profile. This reference profile is the calibrated pulse profile of the same pulsar at $4.8$ GHz provided by the European Pulsar Network Database (\texttt{epndb}) \footnote{\url{http://www.epta.eu.org/epndb/\#hx97b/J0358+5413/hx97b_4850.epn}}. 

\par Bursts from this source are known to have a rotation measure (RM) of $\sim -114$ rad m$^{-2}$ (\citet{20R3Period}, \citetalias{21ZiggyR3,21PastorR3}).
\comment{More recently, it was seen that RM has entered a time variable epoch \citep{R3RM22}. However, for this study a varying low RM does not produce sufficient rotation for bursts with low bandwidth at high frequencies to be fitted for, so we do not fit for RM.} Instead, the calibrated archives are RM corrected using \texttt{pam} \citep{10wvs_pac} with a RM value of $-114.78$ rad m$^{-2}$ \citepalias{21ZiggyR3}. 

\par Calibrated profiles and polarization position angle (PPA) curves of all of our detected bursts are plotted in Fig.~\ref{fig:bursts}. Linear polarization ($L$) was estimated using the unbiased estimator as described in \citet{21NimmoM81,21MicroR3} and \citet{Everett01}. The PPA and its error was also estimated as described in those papers. 
Firstly, the off-pulse statistics was normalized by dividing by the off-pulse standard deviations for each Stokes time series. Linear polarization $L$ was computed as $\sqrt{Q^2 + U^2}$. 
$L$ is de-biased using the off-pulse Stokes-I standard deviation (see Sec.~3.2.1 of \citet{Everett01}). 
PPA errors are estimated using the procedure prescribed in Sec.~3.2.2 of \citet{Everett01}. 

\par Our detected bursts all have $\sim 100\%$ linear polarization fraction with observed fluctuations due to low S/N and low time resolution. 
\citet{22FengSRM}~measures the RM scattering ($\sigma_{RM}$) of \frb. Their predicted polarization fraction at $6$ GHz is $\sim 99.9\%$, in agreement with the observed bursts.

\par Flat polarization position angles (PPA) are seen in burst-\texttt{B} and to a certain extent in bursts-\texttt{F,H}, while the rest of the bursts are less than two samples wide. The flatness of PPAs has already been reported at lower frequencies for the same source \citepalias{21ZiggyR3,21PastorR3} and for other sources such as FRB~20201124A~\citep{21HenningR67}, FRB~20200120E~\citep{21NimmoM81} and FRB~20121102A~\citep{18GajjarR1,Hilmarsson2021}. 

\begin{figure}
    \centering
    \includegraphics[width=0.5\textwidth,keepaspectratio]{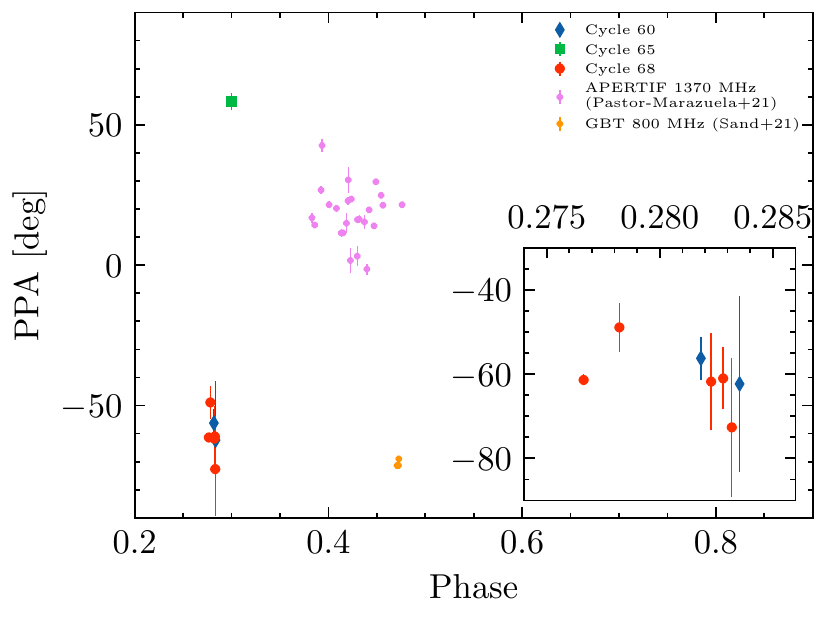}
    \caption{Burst mean Polarization Position Angle (PPA) versus Phase. Mean PPA is the average of the PPA over the ON region. Phase is computed using the periodicity model. Inset zooms in on the points from Cycles 60 and 68 (this work). Note that no absolute PPA angle correction has been done between points from different observatories. }
    \label{fig:paphase}
\end{figure}

In addition, PPAs of all the bursts are plotted against detection phase in  Fig.~\ref{fig:paphase} (also see Tab.~\ref{tab:bursts}). Almost all the bursts having the same PPA except burst-\texttt{C} which has a similar absolute value but with different sign. 
The calibrated test pulsar scan from this day was compared against the reference profile and found to be agreeable. This means that different PPA of burst-\texttt{C} is physical.
\comment{It is interesting to note that multiple bursts detected on two separate epochs (Cycles 60, 68) have consistent PPA value and are within $0.01$ phase units ($\sim 4$ hour duration) around phase $0.28$. However, burst-\texttt{C} which is at phase $0.3$ ($\sim 8$ hours later than the rest.) has a different PPA value.}
Unfortunately, given that it was the only burst detected in that observation and at later phases, we refrain from making any interpretations.

\subsection{Scintillation bandwidth} \label{ssec:scint}

\par \citet{20Marcote} reports a scintillation bandwidth of $59 \pm 13$ kHz at $1.7$ GHz. Scaling it to $4.5$ GHz using typical scaling law of $\sim f^{4.4}$ predicts the scintillation bandwidth to be $\sim 4$ MHz. The channelization of the backend for these observations is $0.97$ MHz and is sufficient to resolve the scintillation bandwidth. Hence, measuring the scintillation bandwidth is attempted here.

\par \comment{To measure the scintillation bandwidths, we follow the procedure of \citet{main+2021}. First, the frequency extent of each burst is identified. Then, the burst flux (which we get after subtracting \texttt{ON-OFF}) over frequency is smoothed by dividing it by a smoothened version of itself. This step normalizes the bandshape. Since the frequency extent of the bursts is about $200$ MHz (c.f. Tab.~\ref{tab:bursts}), a Gaussian kernel with a bandwidth of $200$ channels ($\sim 195$ MHz) is used to compute the smoothed version. Then, autocorrelation functions (ACFs) of the bursts are computed, and using them, scintillation bandwidth is estimated by fitting Lorentzians. While fitting the ACFs, care has been taken to ignore the zero-lag term since it contains the noise correlations.}


\begin{figure*}
\centering
\includegraphics[width=0.95\textwidth,keepaspectratio]{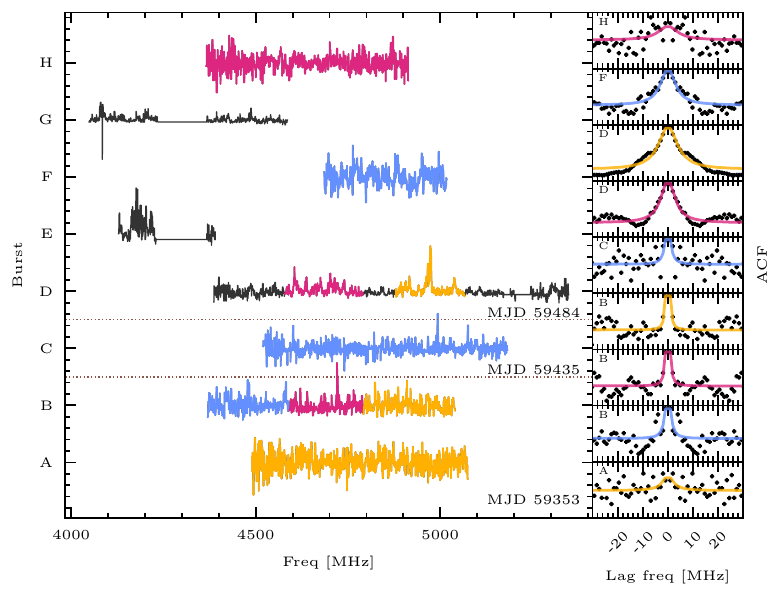}
\caption{Burst spectra and Autocorrelation Functions (ACFs). \emph{Left} panel shows the frequency profile after normalizing (see Sec.~\ref{ssec:scint}). Panels to the \emph{Right} show each individual ACFs of each band/sub-band as scatter plot, and corresponding Lorentzian fit as the solid line. Each band/sub-band are color coded. Scintillation bandwidth is measured by the Lorentzian fits. It is not measured for bursts-\texttt{E,G}, but they are plotted here for completeness. Bursts-\texttt{B,D} are manually sub-banded and each sub-band is used for analysis separately. The rest of the bursts are used in whole. }
\label{fig:burstacf}
\end{figure*}

\par For bursts-\texttt{A,C,F,H}, where the scintles are not so prominent, the full visible band is used for the ACF computation. For bursts-\texttt{B,D}, where the scintles are apparent, sub-bands are chosen manually which cover the scintle feature for the computation. Bursts-\texttt{E,G} are excluded from this analysis due to low S/N per channel and high amount of RFI. 
The frequency of the measured scintillation bandwidth is simply the median of the band/sub-band chosen.
Fig.~\ref{fig:burstacf} illustrates the scintillation spectra and ACFs of the bursts reported here. 

\begin{figure}
    \centering
    \includegraphics[width=0.5\textwidth,keepaspectratio]{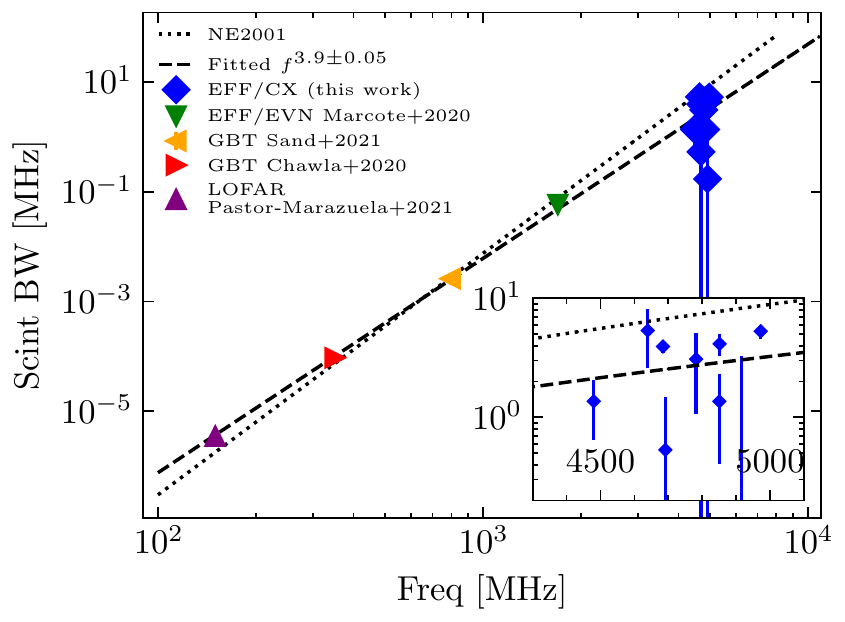}
    \caption{Scintillation bandwidth-frequency dependency. \emph{Inset} zooms in high-frequency detections presented here. The power law fitted using all the reported detections so far. See Sec.~\ref{ssec:scint}. NE2001 \citep{ne2001} is plotted in dotted line. Fitted power law is plotted in broken line.}
    \label{fig:scbwfreq}
\end{figure}

\par There have been multiple measurements of the scattering tail and the scintillation bandwidth at lower frequencies.
\citetalias{21PastorR3} report a scattering timescale of $45$ ms at $150$ MHz, and \citet{20Chawla}~report a scattering upper limit  of $1.7$ ms at $350$ MHz. Both of these numbers can be translated into scintillation bandwidths 
using the relation $\tau_{s} \approx 1/2\pi\nu_{\rm scint}$. \citet{21SandR3}~reports scintillation bandwidths from their GBT detections at $800$ MHz. Lastly, as previously noted in Sec.~\ref{ssec:scint}, \citet{20Marcote}~reports the bandwidth at $1.7$ GHz. This work reports the scintillation bandwidth at $4.5$ GHz which when fitted with other detections yields a power law as $f^{3.90 \pm 0.05}$. This is presented graphically in Fig.~\ref{fig:scbwfreq}. The agreement with the thin scattering screen which requires the frequency dependency as $f^{4}$ is noted. In addition, scintillation bandwidth due to Milky Way is predicted using the NE2001 model \citep{ne2001} and plotted in the same Fig.~\ref{fig:scbwfreq}.  

\par
\begin{figure}
    \centering
    \includegraphics[width=0.5\textwidth,keepaspectratio]{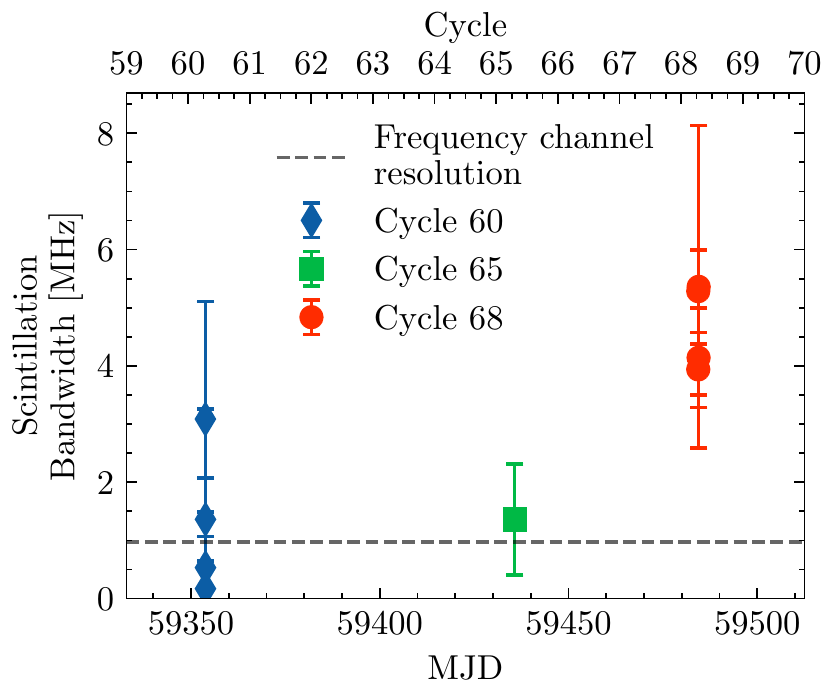}
    \caption{Scintillation bandwidth versus MJD. Top x-axis shows the cycle. All errors are at $1\sigma$. The horizontal broken line is the frequency resolution of the filterbank data.}
    \label{fig:scbwmjd}
\end{figure}

 We report mean scintillation bandwidths at different MJDs scaled to $4.5$ GHz using the $f^{3.9}$ relation derived. The scintillation bandwidths are $1.07 \pm 0.46$ MHz on MJD~59353 (Burst-\texttt{A,B}, Cycle 60),  $1.01 \pm 0.71$ MHz for burst-\texttt{C} observed on MJD~59435 (Cycle 65), and $3.40\pm0.24$ MHz on MJD~59484 (Cycle 68) using Burst-\texttt{D,F,H}.
Between Cycle 60 and Cycle 68, the mean scintillation bandwidth varies by $\sim 4.5\sigma$. Individual scintillation bandwidth measurements are plotted in Fig.~\ref{fig:scbwmjd}.

\subsection{Flux and fluence} \label{ssec:energy}

\par To perform flux calibration,
a System Equivalent Flux Density (SEFD) of $25$ Jy is used\footnote{\url{https://eff100mwiki.mpifr-bonn.mpg.de/doku.php?id=information_for_astronomers:rx:s45mm}}, and an off-pulse region is carefully chosen to convert from data units into flux units. The off-pulse mean is subtracted and the resulting filterbank is converted from S/N units to flux density units using the radiometer equation.
The $2$ GHz band is used to generate time series in flux units, even though the frequency extent of the bursts is much smaller. Fluence is measured as average of the ON-pulse region of the calibrated time series multiplied with the width of the ON-pulse region.
The ON-pulse region is identified visually and used as width since majority of the bursts are not time resolved.
Similarly, the frequency extent of each of the burst is also visually determined.
Peak flux (Jy), fluence (Jy ms), widths (ms), frequency extent (MHz) are tabulated in Tab.~\ref{tab:bursts}.

\subsection{Burst Rate} \label{ssec:rate}

\par The burst rate varies with fluence and is generally reported as the total burst rate above a fluence threshold value.
Firstly, all the published average rates during known active windows at different frequencies are listed below.
\citetalias{21PastorR3} reports rates with a fluence ($\mathcal{F}$) threshold of $50$ Jy ms of $(3.9 \pm 1.3)\times10^{-2}$ \ru at $150$ MHz and $(8.0\pm 1.1)\times10^{-2}$ 
\ru at $1370$ MHz. 
\citetalias{21ZiggyR3} publishes a burst rate of $0.32 \pm 0.08$ \ru above $26$ Jy ms at $150$ MHz and a CHIME/FRB rate of $0.8 \pm 0.3$ \ru at a threshold of $5.1$ Jy ms at $600$ MHz.
Rate scales with this threshold fluence value as $R (\geq \mathcal{F}) \propto \mathcal{F}^{\Gamma}$ \citepalias[c.f.]{21PastorR3} where $\Gamma$ is found to be $-1.5$ \citepalias{21PastorR3}.
It is also noted that the rate varies over phase in its active window \citepalias[See Fig.4]{21PastorR3}. 

\par Then, the average rate during active window ($R$) at $6$ GHz is predicted using the published rate estimates at lower frequencies.
All of the rates are scaled to bring them to the same fluence threshold of $0.1$ Jy ms using the rate-fluence scaling relation mentioned above.
Since \citetalias{21PastorR3} and \citetalias{21ZiggyR3}~both measure a burst rate at $150$ MHz, the mean of both of the rates after being brought to a same fluence threshold used hereon. 
Thereafter, a powerlaw is fitted and extrapolated to frequency of interest. 
Burst rate at $6$ GHz is predicted to be $0.05$ hr$^{-1}$ at threshold fluence of $0.1$ Jy ms.

\par The burst rate at 6 GHz is estimated assuming a fiducial burst with width $1$ ms. This burst would then have the fluence limit of $0.207$ Jy ms (computed using the  quoted SEFD, see Sec.~\ref{ssec:energy}). 
The fluence width scaling is $F = 0.207 \sqrt{ \frac{W}{1 {\rm\ ms}} }$ Jy ms. 
Only three bursts (\texttt{B,D,G}) are above the fluence limit, therefore,
$R (\geq 0.207 {\rm\ Jy\ ms}) = 0.068^{+0.131}_{-0.054} {\rm\ hr}^{-1}$. 
Furthermore, given that most of the bursts are unresolved, yet another fiducial burst with width of one time sample is considered. One time sample burst (of width $0.131$ ms) corresponds to a fluence limit of $0.075$ Jy ms, and the rate is  
$R (\geq 0.075 {\rm\ Jy\ ms}) = 0.18^{+0.18}_{-0.10} {\rm\ hr}^{-1}$.
Note that due to the poor time resolution, the rate reported here is only a lower limit. This could mean that the observed rate is lower than the inferred rate.

\par The measured rate is used to fit a power law%
. The observed rate scales as $f^{-2.64 \pm 0.20}$. The measured rate equation is written in Tab.~\ref{tab:pl}. 
However, we caution that the rate reported here might be underestimated due to the poor time resolution, and the measured rate-frequency relation does not take into account the bursts' limited bandwidth. Modeling such a relation would help greatly in estimating the rate when observing the source with a large bandwidth instrument, and is left for future work.

\par The rate at 6 GHz when measured from each observing session is consistent with rate measured from all the sessions with detections. We report a burst rate of $0.38^{+1.00}_{-0.34}$ \ru in Cycle 60, $0.30^{+1.38}_{-0.29}$ \ru in Cycle 65, and $1.25^{+1.67}_{-0.85}$ \ru in Cycle 68. Rates measured from Cycle 60 and Cycle 65 are consistent with the average rate estimate. However, rate from Cycle 68 is higher where we detected five bursts, suggesting that the source is not well described by a constant Poisson rate over multiple cycles.  
Such a long term non-Poissonianity has already been reported for \rAO{} in \citet{Oppermann2018,Cruces+2021,Li+2021,22JahnsR1}. 
While \rss{} appears to have an underlying Poisson distribution in short observations \citep{22MarthiR67}, it remains to be seen if it behaves differently across multiple observing epochs.
All the errors reported above are at $95\%$ confidence limits. 

\section{Discussion} \label{sec:dis}

\subsection{Evidence for chromaticity at high frequencies} \label{ssec:chf}

\par In this paper, we presented results from two observing campaigns: \psv~performed throughout one activity period with roughly daily cadence and \ptt~in a predicted active window (see Sec.~\ref{ssec:chrom}). 
While both campaigns have similar total exposures (see Tab.~\ref{tab:obs}), \ptt~detected eight bursts, whereas \psv~did not detect any. 

\par An argument for chromaticity at high frequencies is presented as follows:
using the rate estimate computed in Sec.~\ref{ssec:rate} ($0.18$ \ru) from \ptt~assuming Poisson statistics, the probability of detecting at least one burst from \psv~(with exposure of $\sim 35$ hr) is computed under the assumption that the rate estimated in \ptt~holds for the full cycle. 
The expected number of bursts is $\lambda = {\rm Rate} \times {\rm\ exposure} = 6.3$, 
so the probability of detecting at least one burst in \psv~is $99.8\%$. 
On this basis, the hypothesis that the rate is constant over the entire period (which would mean there is no chromaticity at 6 GHz) is rejected with a significance of $\sim 3.11\sigma$.

\par However, there exists a major caveat that has to be mentioned: the cycle-to-cycle variations in rate.
In Sec.~\ref{ssec:rate}, it is shown that the cycle-to-cyle variations in rate are inconsistent with a single Poisson rate.  
So, it could have been that the cycle observed during \psv~had a particularly low burst rate.
Although this is alleviated somewhat by the fact that the average rate used in the above testing is the rate computed across several cycles.

\par Observations in this work were scheduled using a model as described in Sec.~\ref{ssec:obs}. 
Re-fitting the chromatic model with the detected bursts would bias the model and be circular.
Future studies can attempt to improve the fitting with more exposures in and around the predicted window, and with more detections.
In addition, future work should also consider incorporating the chromaticity in a periodic model instead of using a power-law-like relation.  Furthermore, incorporating the chromaticity in physical models would be even more insightful. 

\subsection{High frequency detections}

\par Searches for bursts at higher frequencies help constrain the emission mechanism. To date, high frequency bursts have been detected from \rAO~ \citep{18GajjarR1,Spitler+18,20PearlmanDSN}, FRB~20200120E~\citep{21MajidM81} and FRB~20190520~\citep{AnnaThomas+22}. With this work, \frb~has also joined the rank. Naturally, this begets the question of what is the highest frequency at which bursts can be detectable. 

\par Observations conducted here were in $4-8$ GHz, however bursts were detected only below $5.4$ GHz (see Fig.~\ref{fig:scbwfreq}). This is despite conducting observations in a predicted  window encompassing $5-7$ GHz (c.f. Sec.~\ref{ssec:obs}, Fig.~\ref{fig:mjdp}). 
Non-detection implies several possibilities: the time resolution of the data is not sufficient, the sensitivity of the instrument at higher end of the band lower than expected, or the choice of scan window favored lower end of the band, in which case it would mean that the model is incorrect. The possibility of a cut-off frequency of the bursts cannot be ruled out since this study was done at a poor time resolution. The majority of the bursts detected are not time resolved, which could mean that bursts at higher frequencies are even narrower and the limited time resolution reduced their sensitivity below the detection threshold.
In case the model is incorrect, it would require observations over a wider and likely earlier  phase range to detect a higher frequency burst.

\subsection{Burst properties} \label{ssec:bp}
\begin{figure}
    \centering
    \includegraphics[width=0.49\textwidth,keepaspectratio]{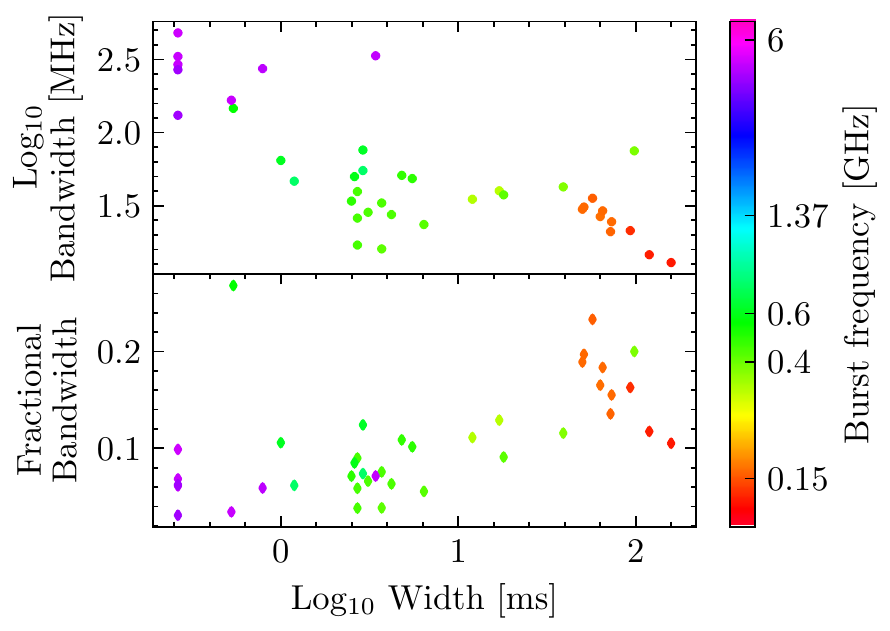}
    \caption{\emph{Top}: Temporal widths versus frequency extent of bursts from \frb. \emph{Bottom}: Temporal widths versus fractional bandwidth. Burst data are taken from \citetalias{21ZiggyR3}, \citetalias{21PastorR3}, \citet{marthi+20} and this work using LOFAR, uGMRT, CHIME/FRB, APERTIF, \eff. All points are color coded by the center frequency of the burst.}
    \label{fig:widths}
\end{figure}
\par Bursts at 150 MHz possess envelope widths of $\sim 50$ ms \citepalias{21ZiggyR3, 21PastorR3}, $\sim 1-2$ ms at 600 MHz \citepalias{21ZiggyR3} and, by this work, $\lesssim 0.1$ ms at 6 GHz (although the bursts at 150 MHz are heavily scatter-broadened \citealt{21PastorR3}). While in \citet{marthi+20}, some bursts appear much wider than 2 ms at $650$ MHz, we note that bursts from \frb{} possess complex morphology. 
Like some giant pulses of pulsars \citep{21Thulasiram,21Geyer} and repeating FRBs (for example, see \citet{21HenningR67}~for FRB~20201124A and \citet{22JahnsR1}~for \rAO), bursts of \frb{} are narrowband. Bursts at any observing frequency has a frequency extent of couple of hundred MHz.
A trend can be seen here: with the temporal widths decreasing, the spectral widths are increasing and the fractional bandwidths are decreasing.
To illustrate this, temporal and spectral widths of the bursts from this work,  \citetalias{21ZiggyR3}, \citetalias{21PastorR3} and \citet{marthi+20} are plotted in the top panel of Fig.~\ref{fig:widths}. The fractional bandwidth is plotted against temporal widths in the bottom panel. Each point is colour coded by the center frequency of the burst. Note that bursts from APERTIF \citepalias{21PastorR3} are not included since all those bursts are instrument bandwidth limited.  
Bursts at lower end of the spectrum ($\sim 150$ MHz) possess larger temporal widths but smaller spectral widths and larger fractional bandwidth. 
However, bursts at the opposite end ($\sim 6$ GHz) possess smaller temporal widths, larger spectral widths and smaller fractional bandwidth. 
The effect of burst widths decreasing at higher frequencies is a well established prediction \citep{78Cordes,92Phillips}. 
However, it is not obvious if burst bandwidth should increase at higher frequencies. 
\comment{In case of \rAO{}, we see a similar trend: \citet{22JahnsR1}~observed bursts at 1.4 GHz which possess mean width of 4 ms and burst bandwidth of 250 MHz, \citet{17LawR1}~reports bursts at 3 GHz with mean widths of $\sim 2$ ms and mean burst bandwidth of $\sim 450$ MHz, and lastly, \citet{18GajjarR1}~reports bursts at 6 GHz possessing $\sim 0.65$ ms widths and $\sim 1.5$ GHz of burst bandwidth.
This might be a quirk of the underlying emission mechanism itself and could be used to constraint such models. 
Lastly, we note that the fractional bandwidth is not constant which implies the magnification of the plasma lensing is not constant. This might question if there is plasma lensing occuring \citep[][c.f.]{18MainNature}.
}


\par Spectral index (SI) is defined as $\alpha$ where the flux densities at different frequencies ($S_f$) follow a power law like $S_f \sim f^\alpha$.
\comment{Instead of $S_f$, it is proposed to consider $\mathcal{F} R_\mathcal{F} (f)$ as flux density, where $\mathcal{F}$ denotes the fluence and $R_\mathcal{F} (f)$ is the rate measured at frequency $f$ above the fluence threshold of $\mathcal{F}$ (see Sec.~\ref{ssec:rate}). }
Dimensionally, both are in the units of Jansky. 
Following \citet[Eq. 2]{houben2019}, the statistical SI can be measured as ratio of rate-frequency index ($\alpha$) and $\Gamma+1$ where $\Gamma$ is the index of the luminosity function \comment{(c.f. \citetalias{21PastorR3}, also see Sec.~\ref{ssec:rate})}.
Note that \citet{houben2019,21SandR3} follow the convention of Rate $\propto \mathcal{F}^{\gamma +1}$ whereas \citetalias{21PastorR3} follows Rate $\propto \mathcal{F}^{\Gamma}$ which is followed here. 
We compute a spectral index of $-1.04 \pm 0.08$, assuming that $\Gamma=-1.5$ \citepalias{21PastorR3}.
\citet{21SandR3}~also measures statistical spectral index in a similar method to be $-0.6^{+1.8}_{-0.9}$ for the same source, which is consistent with our measurement. 
\citetalias{21PastorR3}~reported that activity at 150 MHz is higher than that at 1370 MHz. The negative spectral index measurement qualitatively agrees to this.

\par In case of pulsars, the distribution of $\alpha$ is measured to be a log-normal distribution with mean $\sim~-1.5$ \citep{Bates+13,Jankowski+18}. 
Giant pulses from Crab exhibit a wide spread in the spectral index ranging from -10 to 5 \citep{Ramesh+10,Eftekhari_2016,21Thulasiram}.
A spectral index of $\sim -1.04$ is not implausible in the case of magnetars and pulsars. 
\citet{Lazaridis,Maan+22}~show the time spectral index variability of XTE 1810-197 ranges from $-2$ to $5$.

\par The measured scintillation bandwidth over time is plotted in Fig.~\ref{fig:scbwmjd}, showing a significant variation in the most recent detections.
Given that the scattering is happening in the MW, using \texttt{NE2001} \citep{ne2001}, the probable distance of the scattering screen from the Earth along the LOS of this FRB is estimated as $d_l \sim 2.5$ kpc.
The angular extent of the screen is $\theta = \sqrt{\frac{8{\rm\ ln}(2)c\tau}{d_{\rm eff}}}$ \citep[c.f. Eq. 2]{main+2021}, 
where $\tau \approx 1/2\pi\nu_{\rm s}$ is the scattering timescale which we estimate from the scintillation bandwidth measurement of $\sim 3$\,MHz at 6\,GHz, $c$ is the speed of light in vacuum, and ln is the natural logarithmic function.  
\comment{$d_{\rm eff} \equiv \frac{d d_l}{d-d_l}$ is the effective distance in the FRB-screen-Earth model, where we use $d_{\rm eff} \approx d_l$ since the FRB is extragalactic.}
Assuming the relative velocity between the screen and the Earth to be $v_{\rm le} \sim 30$ km s$^{-1}$ (the orbital velocity of the Earth, assuming no screen velocity), the time taken to see a new portion of the screen is $t \sim \frac{\theta d_l}{v_{le}} \sim 19.2$ days, increasing with $\sim f^{-2}$. Our measured variability of $\nu_{\rm scint}$ is plausible, while changes in the MW scintillation would be expected on timescales of $\gtrsim$ year at L-band and below.

\subsection{Current progenitor models} \label{ssec:cmodels}

\par A key observation of chromaticity (c.f. Fig.~\ref{fig:chrom}, Sec.~\ref{ssec:chrom}) which is to be explained by models is that bursts at higher frequencies occur at earlier phase compared to lower frequencies.
The observation that active windows at higher frequencies are narrower compared to lower frequencies is also suggested by the fitted chromatic model. However, verifying this requires further observations with exposure window covering wider phase range.

\citet{21DongziPA}~(hereafter DL21) explains the chromatic active window with asymmetric emission around the magnetic pole with frequency-related emission height. 
It leads to conically shaped distinct emission regions for distinct frequency bands. 
\citetalias{21DongziPA}~also discusses the PPA behavior of various magnetar models, which can be compared with Figure~\ref{fig:paphase}, and will be discussed in the later part of this section.

\par \citet{BeXRB}~proposes a neutron star in an elliptical orbit with a Be-type star with an accretion disk.
\citet{21WadaBinary}~and \citet{20IokaComb}~also propose the binary-comb model wherein the highly magnetized neutron star is in an elliptical orbit with a companion (either massive star, intermediate mass black hole, or super massive black hole) possessing a wind.
While the burst producing mechanism is different among them, one way 
these models explain chromaticity is by employing free-free absorption in the swept-back wind of the companion. 
The optical depth along the Line-of-Sight from the neutron star to the observer depends on the phase of the neutron star in the orbit.
This dependency, coupled with observatory sensitivity at different bands leads to chromaticity. 

\par \citet{BeXRB}~proposes that bursts are only produced when the neutron star is in the Be-type star disk. This means that while active windows at different frequencies can start at different phases, they must end at the same phase (when the neutron star exits the disk). 
\comment{This is in contradiction with the finding that the active window at 1.4~GHz ends much earlier than the 150~MHz window \citepalias{21PastorR3}.
Within our study at 6 GHz, this would mean activity does not cease beyond the predicted window (c.f. Fig.~\ref{fig:chrom}). Then, explaining non detection of \psv~which was spread across a full period would be difficult. Nevertheless, showing that source does not exhibit any activity beyond predicted windows is an extremely difficult undertaking and it could be that the particular cycle \psv~covered was a dormant one for the source.
}

\par \citet{21WadaBinary}~circumvents this discrepancy by employing the senstivity argument.
Furthermore, it also proposes another scenario wherein it makes use of the binary orbit characteristics to map phase of orbit to the emission frequency of the bursts. When the neutron star is closer to the companion, high frequency bursts are detected. And when it is further away, low frequency bursts are detected.
This scenario can also explain the observed chromaticity. 

\par Slow rotating neutron star model for \frb~is discussed in \citet{20PazULPM} and \citet{Kun_Do}. In the model, the PPA can change as a function of the rotation phase \citepalias{21DongziPA}.
As seen in Fig.~\ref{fig:paphase}, all bursts expect Burst-\texttt{C} are detected at close phase and share similar PPAs.
Burst-\texttt{C} is observed at a later phase than the other 6~GHz detection. There may be an orthogonal PPA transition around phase 0.3. The observed PPA show strong phase-related change, which is still consistent with the slow rotating neutron star model. More detection around phase 0.3 to see whether the PPA jump always happens in a narrow phase window can potentially provide definitive evidence for the model.

\par \citet{22Wei_AllPre}~describes all the precession models which can be divided into two classes:
free and forced precession \citepalias[][see also]{21DongziPA}.
The geometry of forced precession is difficult to introduce non-asymmetric shift of active phase against frequency \citepalias{21DongziPA}, hence are not discussed here.
A hallmark feature of precessing models is that the FRB activity period is the precessional period, and it is time-varying on timescales of $1-100$ years \citep{22Wei_AllPre, 20LevinPrecession, 20ZanazziPrecession}. 
Any change in the period would translate into a proportional change in the lengths of active windows at all frequencies, which leads to changes in the phases of the bursts at some frequency. 
Hence, testing for precessional systems requires modeling a change on activity period over time. 
With continuous monitoring of \frb~with CHIME, detection of a burst at an earlier phase would be strong indicator for period change.
\citetalias{21DongziPA} shows that a free precessing magnetar can explain the observed chromaticity and the PPAs. The clustering seen in PPA versus phase (Fig.~\ref{fig:paphase}) so far agrees with this model.

\section{Conclusions} \label{sec:con}

\begin{itemize}
    \item The chromatic activity window of \frb{} extends to 4$-$8\,GHz 
    (Fig.~\ref{fig:chrom}). A chromatic model was constructed (Tab.~\ref{tab:pl}) using the published detections (Sec.~\ref{ssec:chrom}) at lower frequencies ($\lesssim 2$ GHz). The model was then used predict active windows at higher frequencies ($4-8$ GHz) in which observations were scheduled (Sec.~\ref{ssec:obs}), leading to the detection of eight bursts (Sec.~\ref{sec:dis}, Fig.~\ref{fig:bursts}). Null-detection from an earlier observing campaign was used to provide evidence for chromaticity (Fig.~\ref{fig:mjdp}, Sec.~\ref{ssec:chf}).
    \item Eight bursts were detected showing the following properties:
    \begin{itemize}
        \item Bursts at 6 GHz are found to be narrower in time, wider in frequency and have lower fractional bandwidths compared to lower frequencies bursts (Fig.~\ref{fig:widths}).
        \item Bursts are found to be highly linearly polarized and possess flat Polarization Position Angles (PPAs) (Fig.~\ref{fig:bursts}, Tab.~\ref{tab:bursts}). PPAs of bursts at particular phase are seen to be clustered, however, one of the burst which was detected at different phase and has a different PPA (Fig.~\ref{fig:paphase}).
        \item Scintillation bandwidth scales as $f^{3.90\pm0.05}$ suggesting a thin screen scattering (Fig.~\ref{fig:scbwfreq}).
        Scintillation bandwidth MJD variability is consistent with a screen in the Milky Way (Fig.~\ref{fig:scbwmjd}, Sec.~\ref{ssec:bp}). 
        \item Burst rate varies with frequency as $f^{-2.6\pm0.2}$
        . Rate is seen to vary from cycle to cycle (Sec.~\ref{ssec:rate}). On the basis of rate-frequency relation, spectral index is calculated to $-1.04 \pm 0.08$ (Sec.~\ref{ssec:bp}).
        Poor time resolution of the filterbank data suggests rate may be underestimated (Sec.~\ref{ssec:rate}).
    \end{itemize}
    \item The observed chromaticity is applied to \frb~source models (Sec.~\ref{ssec:cmodels}).
    Observations disfavor the binary model proposed in \citet{BeXRB}. However, binary models described in \citet{21WadaBinary}~can explain the observed chromaticity.
    PPA variations in phase (Fig.~\ref{fig:paphase}) agree with slowly rotating or freely precessing neutron star, in full agreement with the predictions made by \citetalias{21DongziPA}.

\end{itemize}

    In summary, this work reports the first high frequency detection of bursts from \frb{} and establishes that chromaticity exhibited by the source at lower frequencies exists at higher frequencies as well. 
    While the origin of the chromatic periodicity is still unknown, it holds a powerful clue to the physical origin of this source.
    At the time of writing, \rAO{} is the only other FRB repeater with a measured activity period. It would be valuable to determine if it also exhibits similar chromaticity, and if periodicity and chromaticity are common or universal properties of periodically repeating FRBs.
    The bursts detected in this work are undersampled in time - future studies of this and other FRBs at high frequencies will benefit from higher time resolution or baseband data, which will increase sensitivity and the number of burst detections, and allow for a study of burst morphology and polarization.  Future high frequency detections at $4-8$\,GHz and above will better sample and constrain the active window - frequency relation; with improved measurements, physical scenarios can be tested against the chromatic activity cycle, as well as the polarization angle against frequency and activity phase, which may help elucidate the nature of \frb{}.


\section{Acknowledgements}
The authors are thankful of Dr. Alex Kraus for scheduling the observations under tight constraints. Special thanks to Ramesh Karuppusamy for resolving out many issues with the receiver backend. SB is extremely gratituous towards Henning Hilmarrson for help with observing and polarization calibrations.
SB would also like to extend his thanks to Hsiu-Hsien Lin for his comments and Viswesh Marthi for the discussions.
Authors would also like to thank Marilyn Cruces for her comments which improved this work.
We acknowledge use of the CHIME/FRB Public Database, provided at \url{https://www.chime-frb.ca/} by the CHIME/FRB Collaboration. 
Based on observations with the 100-m telescope of the MPIfR (Max-Planck-Institut f{\"u}r Radioastronomie) at Effelsberg.
LGS is a Lise Meitner Indepdendent Max Planck research group leader and acknowledges support from the Max Planck Society. 
Part of this research was carried out at the Jet Propulsion Laboratory, California Institute of Technology, under a contract with the National Aeronautics and Space Administration.

\section{DATA AVAILABILITY}
The data underlying this article will be shared on reasonable request to the corresponding authors. 

\bibliographystyle{mnras}
\bibliography{main}

\begin{thebibliography}{}
\makeatletter
\relax
\def\mn@urlcharsother{\let\do\@makeother \do\$\do\&\do\#\do\^\do\_\do\%\do\~}
\def\mn@doi{\begingroup\mn@urlcharsother \@ifnextchar [ {\mn@doi@}
  {\mn@doi@[]}}
\def\mn@doi@[#1]#2{\def\@tempa{#1}\ifx\@tempa\@empty \href
  {http://dx.doi.org/#2} {doi:#2}\else \href {http://dx.doi.org/#2} {#1}\fi
  \endgroup}
\def\mn@eprint#1#2{\mn@eprint@#1:#2::\@nil}
\def\mn@eprint@arXiv#1{\href {http://arxiv.org/abs/#1} {{\tt arXiv:#1}}}
\def\mn@eprint@dblp#1{\href {http://dblp.uni-trier.de/rec/bibtex/#1.xml}
  {dblp:#1}}
\def\mn@eprint@#1:#2:#3:#4\@nil{\def\@tempa {#1}\def\@tempb {#2}\def\@tempc
  {#3}\ifx \@tempc \@empty \let \@tempc \@tempb \let \@tempb \@tempa \fi \ifx
  \@tempb \@empty \def\@tempb {arXiv}\fi \@ifundefined
  {mn@eprint@\@tempb}{\@tempb:\@tempc}{\expandafter \expandafter \csname
  mn@eprint@\@tempb\endcsname \expandafter{\@tempc}}}

\bibitem[\protect\citeauthoryear{Aggarwal, Law, Burke-Spolaor, Bower, Butler,
  Demorest, Linford  \& Lazio}{Aggarwal et~al.}{2020}]{AtelRF}
Aggarwal K.,  Law C.~J.,  Burke-Spolaor S.,  Bower G.,  Butler B.~J.,  Demorest
  P.,  Linford J.,   Lazio T. J.~W.,  2020, \mn@doi [Research Notes of the
  {AAS}] {10.3847/2515-5172/ab9f33}, 4, 94

\bibitem[\protect\citeauthoryear{{Anna-Thomas} et~al.,}{{Anna-Thomas}
  et~al.}{2022}]{AnnaThomas+22}
{Anna-Thomas} R.,  et~al., 2022, arXiv e-prints, \href
  {https://ui.adsabs.harvard.edu/abs/2022arXiv220211112A} {p. arXiv:2202.11112}

\bibitem[\protect\citeauthoryear{{Bates}, {Lorimer}  \& {Verbiest}}{{Bates}
  et~al.}{2013}]{Bates+13}
{Bates} S.~D.,  {Lorimer} D.~R.,   {Verbiest} J.~P.~W.,  2013, \mn@doi [\mnras]
  {10.1093/mnras/stt257}, \href
  {https://ui.adsabs.harvard.edu/abs/2013MNRAS.431.1352B} {431, 1352}

\bibitem[\protect\citeauthoryear{{Beniamini}, {Wadiasingh}  \&
  {Metzger}}{{Beniamini} et~al.}{2020}]{20PazULPM}
{Beniamini} P.,  {Wadiasingh} Z.,   {Metzger} B.~D.,  2020, \mn@doi [\mnras]
  {10.1093/mnras/staa1783}, \href
  {https://ui.adsabs.harvard.edu/abs/2020MNRAS.496.3390B} {496, 3390}

\bibitem[\protect\citeauthoryear{{Bhardwaj} et~al.,}{{Bhardwaj}
  et~al.}{2021}]{Bhardwaj2021}
{Bhardwaj} M.,  et~al., 2021, \mn@doi [\apjl] {10.3847/2041-8213/abeaa6}, \href
  {https://ui.adsabs.harvard.edu/abs/2021ApJ...910L..18B} {910, L18}

\bibitem[\protect\citeauthoryear{{Chawla} et~al.,}{{Chawla}
  et~al.}{2020}]{20Chawla}
{Chawla} P.,  et~al., 2020, \mn@doi [\apjl] {10.3847/2041-8213/ab96bf}, \href
  {https://ui.adsabs.harvard.edu/abs/2020ApJ...896L..41C} {896, L41}

\bibitem[\protect\citeauthoryear{{Chime/Frb Collaboration} et~al.,}{{Chime/Frb
  Collaboration} et~al.}{2020}]{20R3Period}
{Chime/Frb Collaboration} et~al., 2020, \mn@doi [\nat]
  {10.1038/s41586-020-2398-2}, \href
  {https://ui.adsabs.harvard.edu/abs/2020Natur.582..351C} {582, 351}

\bibitem[\protect\citeauthoryear{{Cordes}}{{Cordes}}{1978}]{78Cordes}
{Cordes} J.~M.,  1978, \mn@doi [\apj] {10.1086/156218}, \href
  {https://ui.adsabs.harvard.edu/abs/1978ApJ...222.1006C} {222, 1006}

\bibitem[\protect\citeauthoryear{{Cordes} \& {Lazio}}{{Cordes} \&
  {Lazio}}{2002}]{ne2001}
{Cordes} J.~M.,  {Lazio} T.~J.~W.,  2002, arXiv e-prints, \href
  {https://ui.adsabs.harvard.edu/abs/2002astro.ph..7156C} {pp
  astro--ph/0207156}

\bibitem[\protect\citeauthoryear{{Cruces} et~al.,}{{Cruces}
  et~al.}{2021}]{Cruces+2021}
{Cruces} M.,  et~al., 2021, \mn@doi [\mnras] {10.1093/mnras/staa3223}, \href
  {https://ui.adsabs.harvard.edu/abs/2021MNRAS.500..448C} {500, 448}

\bibitem[\protect\citeauthoryear{Eftekhari, Stovall, Dowell, Schinzel  \&
  Taylor}{Eftekhari et~al.}{2016}]{Eftekhari_2016}
Eftekhari T.,  Stovall K.,  Dowell J.,  Schinzel F.~K.,   Taylor G.~B.,  2016,
  \mn@doi [The Astrophysical Journal] {10.3847/0004-637x/829/2/62}, 829, 62

\bibitem[\protect\citeauthoryear{{Everett} \& {Weisberg}}{{Everett} \&
  {Weisberg}}{2001}]{Everett01}
{Everett} J.~E.,  {Weisberg} J.~M.,  2001, \mn@doi [\apj] {10.1086/320652},
  \href {https://ui.adsabs.harvard.edu/abs/2001ApJ...553..341E} {553, 341}

\bibitem[\protect\citeauthoryear{{Feng} et~al.,}{{Feng}
  et~al.}{2022}]{22FengSRM}
{Feng} Y.,  et~al., 2022, arXiv e-prints, \href
  {https://ui.adsabs.harvard.edu/abs/2022arXiv220209601F} {p. arXiv:2202.09601}

\bibitem[\protect\citeauthoryear{{Gajjar} et~al.,}{{Gajjar}
  et~al.}{2018}]{18GajjarR1}
{Gajjar} V.,  et~al., 2018, \mn@doi [\apj] {10.3847/1538-4357/aad005}, \href
  {https://ui.adsabs.harvard.edu/abs/2018ApJ...863....2G} {863, 2}

\bibitem[\protect\citeauthoryear{{Geyer} et~al.,}{{Geyer}
  et~al.}{2021}]{21Geyer}
{Geyer} M.,  et~al., 2021, \mn@doi [\mnras] {10.1093/mnras/stab1501}, \href
  {https://ui.adsabs.harvard.edu/abs/2021MNRAS.505.4468G} {505, 4468}

\bibitem[\protect\citeauthoryear{{Hilmarsson}, {Spitler}, {Main}  \&
  {Li}}{{Hilmarsson} et~al.}{2021a}]{21HenningR67}
{Hilmarsson} G.~H.,  {Spitler} L.~G.,  {Main} R.~A.,   {Li} D.~Z.,  2021a,
  \mn@doi [\mnras] {10.1093/mnras/stab2936}, \href
  {https://ui.adsabs.harvard.edu/abs/2021MNRAS.508.5354H} {508, 5354}

\bibitem[\protect\citeauthoryear{{Hilmarsson} et~al.,}{{Hilmarsson}
  et~al.}{2021b}]{Hilmarsson2021}
{Hilmarsson} G.~H.,  et~al., 2021b, \mn@doi [\apjl] {10.3847/2041-8213/abdec0},
  \href {https://ui.adsabs.harvard.edu/abs/2021ApJ...908L..10H} {908, L10}

\bibitem[\protect\citeauthoryear{{Hotan}, {van Straten}  \&
  {Manchester}}{{Hotan} et~al.}{2004}]{psrchive}
{Hotan} A.~W.,  {van Straten} W.,   {Manchester} R.~N.,  2004, \mn@doi [\pasa]
  {10.1071/AS04022}, \href
  {https://ui.adsabs.harvard.edu/abs/2004PASA...21..302H} {21, 302}

\bibitem[\protect\citeauthoryear{{Houben}, {Spitler}, {ter Veen}, {Rachen},
  {Falcke}  \& {Kramer}}{{Houben} et~al.}{2019}]{houben2019}
{Houben} L.~J.~M.,  {Spitler} L.~G.,  {ter Veen} S.,  {Rachen} J.~P.,  {Falcke}
  H.,   {Kramer} M.,  2019, \mn@doi [A\&A] {10.1051/0004-6361/201833875}, \href
  {https://ui.adsabs.harvard.edu/abs/2019A&A...623A..42H} {623, A42}

\bibitem[\protect\citeauthoryear{{Ioka} \& {Zhang}}{{Ioka} \&
  {Zhang}}{2020}]{20IokaComb}
{Ioka} K.,  {Zhang} B.,  2020, \mn@doi [\apjl] {10.3847/2041-8213/ab83fb},
  \href {https://ui.adsabs.harvard.edu/abs/2020ApJ...893L..26I} {893, L26}

\bibitem[\protect\citeauthoryear{{Jahns} et~al.,}{{Jahns}
  et~al.}{2022}]{22JahnsR1}
{Jahns} J.~N.,  et~al., 2022, arXiv e-prints, \href
  {https://ui.adsabs.harvard.edu/abs/2022arXiv220205705J} {p. arXiv:2202.05705}

\bibitem[\protect\citeauthoryear{{Jankowski}, {van Straten}, {Keane}, {Bailes},
  {Barr}, {Johnston}  \& {Kerr}}{{Jankowski} et~al.}{2018}]{Jankowski+18}
{Jankowski} F.,  {van Straten} W.,  {Keane} E.~F.,  {Bailes} M.,  {Barr} E.~D.,
   {Johnston} S.,   {Kerr} M.,  2018, \mn@doi [\mnras] {10.1093/mnras/stx2476},
  \href {https://ui.adsabs.harvard.edu/abs/2018MNRAS.473.4436J} {473, 4436}

\bibitem[\protect\citeauthoryear{{Karuppusamy}, {Stappers}  \& {van
  Straten}}{{Karuppusamy} et~al.}{2010}]{Ramesh+10}
{Karuppusamy} R.,  {Stappers} B.~W.,   {van Straten} W.,  2010, \mn@doi [\aap]
  {10.1051/0004-6361/200913729}, \href
  {https://ui.adsabs.harvard.edu/abs/2010A&A...515A..36K} {515, A36}

\bibitem[\protect\citeauthoryear{{Law} et~al.,}{{Law} et~al.}{2017}]{17LawR1}
{Law} C.~J.,  et~al., 2017, \mn@doi [\apj] {10.3847/1538-4357/aa9700}, \href
  {https://ui.adsabs.harvard.edu/abs/2017ApJ...850...76L} {850, 76}

\bibitem[\protect\citeauthoryear{{Lazaridis}, {Jessner}, {Kramer}, {Stappers},
  {Lyne}, {Jordan}, {Serylak}  \& {Zensus}}{{Lazaridis}
  et~al.}{2008}]{Lazaridis}
{Lazaridis} K.,  {Jessner} A.,  {Kramer} M.,  {Stappers} B.~W.,  {Lyne} A.~G.,
  {Jordan} C.~A.,  {Serylak} M.,   {Zensus} J.~A.,  2008, \mn@doi [\mnras]
  {10.1111/j.1365-2966.2008.13794.x}, \href
  {https://ui.adsabs.harvard.edu/abs/2008MNRAS.390..839L} {390, 839}

\bibitem[\protect\citeauthoryear{{Levin}, {Beloborodov}  \&
  {Bransgrove}}{{Levin} et~al.}{2020}]{20LevinPrecession}
{Levin} Y.,  {Beloborodov} A.~M.,   {Bransgrove} A.,  2020, \mn@doi [\apjl]
  {10.3847/2041-8213/ab8c4c}, \href
  {https://ui.adsabs.harvard.edu/abs/2020ApJ...895L..30L} {895, L30}

\bibitem[\protect\citeauthoryear{{Li} \& {Zanazzi}}{{Li} \&
  {Zanazzi}}{2021}]{21DongziPA}
{Li} D.,  {Zanazzi} J.~J.,  2021, \mn@doi [\apjl] {10.3847/2041-8213/abeaa4},
  \href {https://ui.adsabs.harvard.edu/abs/2021ApJ...909L..25L} {909, L25}

\bibitem[\protect\citeauthoryear{{Li} et~al.,}{{Li} et~al.}{2021a}]{Li+2021}
{Li} D.,  et~al., 2021a, arXiv e-prints, \href
  {https://ui.adsabs.harvard.edu/abs/2021arXiv210708205L} {p. arXiv:2107.08205}

\bibitem[\protect\citeauthoryear{{Li}, {Yang}, {Wang}, {Xu}, {Shao}, {Liu}  \&
  {Dai}}{{Li} et~al.}{2021b}]{BeXRB}
{Li} Q.-C.,  {Yang} Y.-P.,  {Wang} F.~Y.,  {Xu} K.,  {Shao} Y.,  {Liu} Z.-N.,
  {Dai} Z.-G.,  2021b, \mn@doi [\apjl] {10.3847/2041-8213/ac1922}, \href
  {https://ui.adsabs.harvard.edu/abs/2021ApJ...918L...5L} {918, L5}

\bibitem[\protect\citeauthoryear{{Lorimer}, {Bailes}, {McLaughlin}, {Narkevic}
  \& {Crawford}}{{Lorimer} et~al.}{2007}]{lorimer2007}
{Lorimer} D.~R.,  {Bailes} M.,  {McLaughlin} M.~A.,  {Narkevic} D.~J.,
  {Crawford} F.,  2007, \mn@doi [Science] {10.1126/science.1147532}, \href
  {https://ui.adsabs.harvard.edu/abs/2007Sci...318..777L} {318, 777}

\bibitem[\protect\citeauthoryear{{Maan}, {Surnis}, {Joshi}  \& {Bagchi}}{{Maan}
  et~al.}{2022}]{Maan+22}
{Maan} Y.,  {Surnis} M.~P.,  {Joshi} B.~C.,   {Bagchi} M.,  2022, arXiv
  e-prints, \href {https://ui.adsabs.harvard.edu/abs/2022arXiv220113006M} {p.
  arXiv:2201.13006}

\bibitem[\protect\citeauthoryear{{Main} et~al.,}{{Main}
  et~al.}{2018}]{18MainNature}
{Main} R.,  et~al., 2018, \mn@doi [\nat] {10.1038/s41586-018-0133-z}, \href
  {https://ui.adsabs.harvard.edu/abs/2018Natur.557..522M} {557, 522}

\bibitem[\protect\citeauthoryear{{Main}, {Hilmarsson}, {Marthi}, {Spitler},
  {Wharton}, {Bethapudi}, {Li}  \& {Lin}}{{Main} et~al.}{2022}]{main+2021}
{Main} R.~A.,  {Hilmarsson} G.~H.,  {Marthi} V.~R.,  {Spitler} L.~G.,
  {Wharton} R.~S.,  {Bethapudi} S.,  {Li} D.~Z.,   {Lin} H.~H.,  2022, \mn@doi
  [\mnras] {10.1093/mnras/stab3218}, \href
  {https://ui.adsabs.harvard.edu/abs/2022MNRAS.509.3172M} {509, 3172}

\bibitem[\protect\citeauthoryear{{Majid} et~al.,}{{Majid}
  et~al.}{2021}]{21MajidM81}
{Majid} W.~A.,  et~al., 2021, \mn@doi [\apjl] {10.3847/2041-8213/ac1921}, \href
  {https://ui.adsabs.harvard.edu/abs/2021ApJ...919L...6M} {919, L6}

\bibitem[\protect\citeauthoryear{{Mannings} et~al.,}{{Mannings}
  et~al.}{2021}]{mannings+21}
{Mannings} A.~G.,  et~al., 2021, \mn@doi [\apj] {10.3847/1538-4357/abff56},
  \href {https://ui.adsabs.harvard.edu/abs/2021ApJ...917...75M} {917, 75}

\bibitem[\protect\citeauthoryear{{Marcote} et~al.,}{{Marcote}
  et~al.}{2020}]{20Marcote}
{Marcote} B.,  et~al., 2020, \mn@doi [\nat] {10.1038/s41586-019-1866-z}, \href
  {https://ui.adsabs.harvard.edu/abs/2020Natur.577..190M} {577, 190}

\bibitem[\protect\citeauthoryear{{Marthi}, {Gautam}, {Li}, {Lin}, {Main},
  {Naidu}, {Pen}  \& {Wharton}}{{Marthi} et~al.}{2020}]{marthi+20}
{Marthi} V.~R.,  {Gautam} T.,  {Li} D.~Z.,  {Lin} H.~H.,  {Main} R.~A.,
  {Naidu} A.,  {Pen} U.~L.,   {Wharton} R.~S.,  2020, \mn@doi [\mnras]
  {10.1093/mnrasl/slaa148}, \href
  {https://ui.adsabs.harvard.edu/abs/2020MNRAS.499L..16M} {499, L16}

\bibitem[\protect\citeauthoryear{{Marthi} et~al.,}{{Marthi}
  et~al.}{2022}]{22MarthiR67}
{Marthi} V.~R.,  et~al., 2022, \mn@doi [\mnras] {10.1093/mnras/stab3067}, \href
  {https://ui.adsabs.harvard.edu/abs/2022MNRAS.509.2209M} {509, 2209}

\bibitem[\protect\citeauthoryear{{Mckinven} et~al.,}{{Mckinven}
  et~al.}{2022}]{R3RM22}
{Mckinven} R.,  et~al., 2022, arXiv e-prints, \href
  {https://ui.adsabs.harvard.edu/abs/2022arXiv220509221M} {p. arXiv:2205.09221}

\bibitem[\protect\citeauthoryear{{Nimmo} et~al.,}{{Nimmo}
  et~al.}{2021}]{21MicroR3}
{Nimmo} K.,  et~al., 2021, \mn@doi [Nature Astronomy]
  {10.1038/s41550-021-01321-3}, \href
  {https://ui.adsabs.harvard.edu/abs/2021NatAs...5..594N} {5, 594}

\bibitem[\protect\citeauthoryear{{Nimmo} et~al.,}{{Nimmo}
  et~al.}{2022}]{21NimmoM81}
{Nimmo} K.,  et~al., 2022, \mn@doi [Nature Astronomy]
  {10.1038/s41550-021-01569-9}, \href
  {https://ui.adsabs.harvard.edu/abs/2022NatAs...6..393N} {6, 393}

\bibitem[\protect\citeauthoryear{{Niu} et~al.,}{{Niu} et~al.}{2021}]{Niu+21}
{Niu} C.~H.,  et~al., 2021, arXiv e-prints, \href
  {https://ui.adsabs.harvard.edu/abs/2021arXiv211007418N} {p. arXiv:2110.07418}

\bibitem[\protect\citeauthoryear{{Oppermann}, {Yu}  \& {Pen}}{{Oppermann}
  et~al.}{2018}]{Oppermann2018}
{Oppermann} N.,  {Yu} H.-R.,   {Pen} U.-L.,  2018, \mn@doi [\mnras]
  {10.1093/mnras/sty004}, \href
  {https://ui.adsabs.harvard.edu/abs/2018MNRAS.475.5109O} {475, 5109}

\bibitem[\protect\citeauthoryear{{Pastor-Marazuela} et~al.,}{{Pastor-Marazuela}
  et~al.}{2021}]{21PastorR3}
{Pastor-Marazuela} I.,  et~al., 2021, \mn@doi [\nat]
  {10.1038/s41586-021-03724-8}, \href
  {https://ui.adsabs.harvard.edu/abs/2021Natur.596..505P} {596, 505}

\bibitem[\protect\citeauthoryear{{Pearlman}, {Majid}, {Prince}, {Nimmo},
  {Hessels}, {Naudet}  \& {Kocz}}{{Pearlman} et~al.}{2020}]{20PearlmanDSN}
{Pearlman} A.~B.,  {Majid} W.~A.,  {Prince} T.~A.,  {Nimmo} K.,  {Hessels} J.
  W.~T.,  {Naudet} C.~J.,   {Kocz} J.,  2020, \mn@doi [\apjl]
  {10.3847/2041-8213/abca31}, \href
  {https://ui.adsabs.harvard.edu/abs/2020ApJ...905L..27P} {905, L27}

\bibitem[\protect\citeauthoryear{{Phillips}}{{Phillips}}{1992}]{92Phillips}
{Phillips} J.~A.,  1992, \mn@doi [\apj] {10.1086/170936}, \href
  {https://ui.adsabs.harvard.edu/abs/1992ApJ...385..282P} {385, 282}

\bibitem[\protect\citeauthoryear{{Pilia} et~al.,}{{Pilia}
  et~al.}{2020}]{20PiliaR3}
{Pilia} M.,  et~al., 2020, \mn@doi [\apjl] {10.3847/2041-8213/ab96c0}, \href
  {https://ui.adsabs.harvard.edu/abs/2020ApJ...896L..40P} {896, L40}

\bibitem[\protect\citeauthoryear{{Pleunis} et~al.,}{{Pleunis}
  et~al.}{2021}]{21ZiggyR3}
{Pleunis} Z.,  et~al., 2021, \mn@doi [\apjl] {10.3847/2041-8213/abec72}, \href
  {https://ui.adsabs.harvard.edu/abs/2021ApJ...911L...3P} {911, L3}

\bibitem[\protect\citeauthoryear{{Ransom}}{{Ransom}}{2011}]{presto}
{Ransom} S.,  2011, {PRESTO: PulsaR Exploration and Search TOolkit} (\mn@eprint
  {ascl} {1107.017})

\bibitem[\protect\citeauthoryear{{Sand} et~al.,}{{Sand}
  et~al.}{2021}]{21SandR3}
{Sand} K.~R.,  et~al., 2021, arXiv e-prints, \href
  {https://ui.adsabs.harvard.edu/abs/2021arXiv211102382S} {p. arXiv:2111.02382}

\bibitem[\protect\citeauthoryear{{Scholz} et~al.,}{{Scholz}
  et~al.}{2020}]{20ScholzR3Xray}
{Scholz} P.,  et~al., 2020, \mn@doi [\apj] {10.3847/1538-4357/abb1a8}, \href
  {https://ui.adsabs.harvard.edu/abs/2020ApJ...901..165S} {901, 165}

\bibitem[\protect\citeauthoryear{{Spitler} et~al.,}{{Spitler}
  et~al.}{2016}]{Spitler2016}
{Spitler} L.~G.,  et~al., 2016, \mn@doi [\nat] {10.1038/nature17168}, \href
  {https://ui.adsabs.harvard.edu/abs/2016Natur.531..202S} {531, 202}

\bibitem[\protect\citeauthoryear{{Spitler} et~al.,}{{Spitler}
  et~al.}{2018}]{Spitler+18}
{Spitler} L.~G.,  et~al., 2018, \mn@doi [\apj] {10.3847/1538-4357/aad332},
  \href {https://ui.adsabs.harvard.edu/abs/2018ApJ...863..150S} {863, 150}

\bibitem[\protect\citeauthoryear{{Tavani} et~al.,}{{Tavani}
  et~al.}{2020}]{20TavaniGX}
{Tavani} M.,  et~al., 2020, \mn@doi [\apjl] {10.3847/2041-8213/ab86b1}, \href
  {https://ui.adsabs.harvard.edu/abs/2020ApJ...893L..42T} {893, L42}

\bibitem[\protect\citeauthoryear{{Tendulkar} et~al.,}{{Tendulkar}
  et~al.}{2021}]{tendulkar2021}
{Tendulkar} S.~P.,  et~al., 2021, \mn@doi [\apjl] {10.3847/2041-8213/abdb38},
  \href {https://ui.adsabs.harvard.edu/abs/2021ApJ...908L..12T} {908, L12}

\bibitem[\protect\citeauthoryear{{The CHIME Collaboration} et~al.,}{{The CHIME
  Collaboration} et~al.}{2022}]{chime}
{The CHIME Collaboration} et~al., 2022, arXiv e-prints, \href
  {https://ui.adsabs.harvard.edu/abs/2022arXiv220107869T} {p. arXiv:2201.07869}

\bibitem[\protect\citeauthoryear{{Thornton} et~al.,}{{Thornton}
  et~al.}{2013}]{Thornton+13}
{Thornton} D.,  et~al., 2013, \mn@doi [Science] {10.1126/science.1236789},
  \href {https://ui.adsabs.harvard.edu/abs/2013Sci...341...53T} {341, 53}

\bibitem[\protect\citeauthoryear{{Thulasiram} \& {Lin}}{{Thulasiram} \&
  {Lin}}{2021}]{21Thulasiram}
{Thulasiram} P.,  {Lin} H.-H.,  2021, \mn@doi [\mnras]
  {10.1093/mnras/stab2692}, \href
  {https://ui.adsabs.harvard.edu/abs/2021MNRAS.508.1947T} {508, 1947}

\bibitem[\protect\citeauthoryear{{Wada}, {Ioka}  \& {Zhang}}{{Wada}
  et~al.}{2021}]{21WadaBinary}
{Wada} T.,  {Ioka} K.,   {Zhang} B.,  2021, \mn@doi [\apj]
  {10.3847/1538-4357/ac127a}, \href
  {https://ui.adsabs.harvard.edu/abs/2021ApJ...920...54W} {920, 54}

\bibitem[\protect\citeauthoryear{{Wei}, {Zhao}  \& {Wang}}{{Wei}
  et~al.}{2022}]{22Wei_AllPre}
{Wei} Y.-J.,  {Zhao} Z.-Y.,   {Wang} F.-Y.,  2022, \mn@doi [\aap]
  {10.1051/0004-6361/202142321}, \href
  {https://ui.adsabs.harvard.edu/abs/2022A&A...658A.163W} {658, A163}

\bibitem[\protect\citeauthoryear{{Xu}, {Li}, {Yang}, {Li}, {Dai}  \&
  {Liu}}{{Xu} et~al.}{2021}]{Kun_Do}
{Xu} K.,  {Li} Q.-C.,  {Yang} Y.-P.,  {Li} X.-D.,  {Dai} Z.-G.,   {Liu} J.,
  2021, \mn@doi [\apj] {10.3847/1538-4357/ac05ba}, \href
  {https://ui.adsabs.harvard.edu/abs/2021ApJ...917....2X} {917, 2}

\bibitem[\protect\citeauthoryear{{Zanazzi} \& {Lai}}{{Zanazzi} \&
  {Lai}}{2020}]{20ZanazziPrecession}
{Zanazzi} J.~J.,  {Lai} D.,  2020, \mn@doi [\apjl] {10.3847/2041-8213/ab7cdd},
  \href {https://ui.adsabs.harvard.edu/abs/2020ApJ...892L..15Z} {892, L15}

\bibitem[\protect\citeauthoryear{{van Straten}, {Manchester}, {Johnston}  \&
  {Reynolds}}{{van Straten} et~al.}{2010}]{10wvs_pac}
{van Straten} W.,  {Manchester} R.~N.,  {Johnston} S.,   {Reynolds} J.~E.,
  2010, \mn@doi [\pasa] {10.1071/AS09084}, \href
  {https://ui.adsabs.harvard.edu/abs/2010PASA...27..104V} {27, 104}

\makeatother
\end{thebibliography}

\end{document}